\documentclass[10pt, conference, letterpaper]{IEEEtran}
\usepackage{amssymb}        % for set of natural numbers
\usepackage[lined,boxed,linesnumbered,ruled,vlined]{algorithm2e}
\SetNlSty{}{}{:}	        % to add colon after line number
\SetKwInOut{Input}{Input}  
\SetKwInOut{Output}{Output}
\SetKwInOut{Returns}{Return}  
\SetKwInOut{Initialize}{Initialize} 
\usepackage[noend]{algorithmic} % added noend
\usepackage{pgf}                % for plots
\usepackage{tabularx}   % to have text wrap; table span single column of page
\usepackage{url}        % to reference webpages; URLs in the references
\usepackage{amssymb} % added for set of natural numbers

\usepackage[lined,boxed,linesnumbered,ruled,vlined]{algorithm2e} % added this package :(
\SetKwInOut{Input}{Input}  % added
\SetKwInOut{Output}{Output}	% added
\SetKwInOut{Returns}{Return}  % added
\SetKwInOut{Initialize}{Initialize} % added this 
\SetKwComment{Comment}{$ \%\ $}{}
\usepackage[noend]{algorithmic} % added noend so that end if, end for are not added

\usepackage[autostyle]{csquotes}

\usepackage{amsmath}

\usepackage{subcaption}
\usepackage{tablefootnote}

\usepackage{dblfloatfix} % to enable figures at the bottom of page
\newtheorem{theorem}{Theorem}
\newtheorem{lemma}{Lemma} % different counter for lemma and theorem

\title{Cooperation Speeds Surfing: Use Co-Bandit!}
\author{\IEEEauthorblockN{Anuja Meetoo Appavoo, Seth Gilbert, and Kian-Lee Tan}
\IEEEauthorblockA{Department of Computer Science, National University of Singapore}
\{anuja, seth.gilbert, tankl\}@comp.nus.edu.sg}
\begin{document}

\maketitle

\begin{abstract} 
In this paper, we explore the benefit of cooperation in adversarial bandit settings. As a motivating example, we consider the problem of wireless network selection. Mobile devices are often required to choose the right network to associate with for optimal performance, which is non-trivial. The excellent theoretical properties of EXP3, a leading multi-armed bandit algorithm, suggest that it should work well for this type of problem. Yet, it performs poorly in practice. A major limitation is its slow rate of stabilization. 
Bandit-style algorithms perform better when global knowledge is available, i.e., when devices receive feedback about all networks after each selection. But, unfortunately, communicating full information to all devices is expensive. Therefore, we address the question of how much information is adequate to achieve better performance.
% Full information setting requires coordination from network service providers which may be infeasible. Thus, we consider a spectrum of settings that bridges the full information and bandit settings, with cooperation among devices to solve a common problem.

We propose Co-Bandit, a novel cooperative bandit approach, that allows devices to occasionally share their observations and forward feedback received from neighbors; hence, feedback may be received with a delay. Devices perform network selection based on their own observation and feedback from neighbors. As such, they speed up each other's rate of learning. We prove that Co-Bandit is regret-minimizing and retains the convergence property of multiplicative weight update algorithms with full information. Through simulation, we show that a very small amount of information, even with a delay, is adequate to nudge each other to select the right network and yield significantly faster stabilization at the optimal state (about 630x faster than EXP3).
% 84608/134.5
\end{abstract}

\section{Introduction} \label{section:introduction}            Mobile devices often have to select the right network to associate with for optimal performance, which is non-trivial. This is primarily because network availability is transient and the quality of networks changes dynamically due to mobility of devices and environmental factors. The challenge is for devices to make decentralized decisions and yet achieve an optimal allocation, where no device would want to unilaterally change network. Moreover, since the environment is dynamic, devices must be able to seamlessly adapt their decisions to maintain a good network connection. Resource selection problems can be formulated as a repeated congestion game; in each round, a device selects a network and receives some reward, i.e., bandwidth. The multi-armed bandit problem is closely related to repeated multi-player games, where each player independently aims at improving its decision while all players collectively act as an adversary. Multi-armed bandit algorithms have impressive theoretical properties which suggest that they provide an excellent solution to this problem. 

EXP3 (Exponential-weight algorithm for Exploration and Exploitation) \cite{auer2002nonstochastic} is one of the leading bandit algorithms. It is fully decentralized, is regret-minimizing, i.e., as time elapses, it performs nearly as well as always selecting the best action in hindsight, and converges to a (weakly stable) Nash equilibrium \cite{kleinberg2009multiplicative, tekin2011performance}. However, it performs poorly in practice. One major limitation is that it takes an unacceptable amount of time to stabilize; it took the equivalent of over 14 days in some of our simulations.
% In our prior work, we proposed Smart EXP3 \cite{appavoo2018shrewd} with significant performance improvement (over 235x faster stabilization).
% 
The availability of global knowledge would yield better performance. But, it requires support from network service providers, which may be infeasible. 
In this paper, we explore the possibility of cooperation among devices, and consider the tradeoff between the amount of cooperation and performance. We show that a very small amount of information yields massive performance improvement, in terms of rate of stabilization to an optimal state. 
% bridge the full information and bandit settings and consider a spectrum of settings in between. 
% We explore the possibility of cooperation among devices and show that it can significantly speed up the rate of stabilization to an optimal state.
Several mobile applications to-date rely on the cooperation of peers \cite{googlemap, waze} which suggests the feasibility of cooperation for wireless network selection.

We formulate the wireless network selection problem as a repeated congestion game, and model the behaviour of mobile devices using online learning with partial information and delayed feedback in an adversarial setting. We propose Co-Bandit, a novel cooperative bandit algorithm with good theoretical and practical performance. Mobile devices (a) occasionally share their observations 
% with their neighbors
(without overloading the network), e.g., they can broadcast bit rates observed from their chosen networks using Bluetooth, (b) forward feedback received over a certain period of time; hence, feedback may be received with a delay, and (c) use feedback received from neighbors to enhance their decisions. We model the underlying communication network as a random directed graph based on communication pattern in a wireless network setting. Vertices represent mobile devices and a directed edge implies sharing of feedback. Moreover, the topology of the graph changes over time. 
% Through simulation, we observe that a very small amount of information, even with a delay, is enough to nudge each other to select the right network. 
All source code is available on GitHub \footnote{https://github.com/anuja-meetoo/Co-Bandit}.

To summarize, the following are our key contributions:
\begin{enumerate}
    \item We formulate the network selection problem as a repeated congestion game and model the behaviour of devices using online learning with partial information and delayed feedback in an adversarial setting.
    \item We propose Co-Bandit, a cooperative bandit approach, that allows devices to share their observations, forward feedback received, and perform network selection based on their observation as well as those received from neighbours, at times with a delay.
    \item We prove that Co-Bandit (a) is regret-minimizing and provide an upper bound that highlights the effect of cooperation and delay, and (b) retains the convergence property of multiplicative weight update algorithms with full information.
    \item Through simulation, we demonstrate that Co-Bandit (a) stabilizes at the optimal state relatively fast with only a small amount of information, (b) gracefully deals with transient behaviours, and (c) scales with an increase in number of devices and/or networks.
\end{enumerate}

\section{Wireless network selection} \label{section:problemFormulation}
    % In this section
Here, we define the wireless network selection problem, formulate it as a repeated congestion game, and model the behavior of mobile devices using online learning with partial information and delayed feedback in an adversarial setting.

\subsection{Wireless network selection problem}
\begin{figure}[!htb]
\begin{center}
% trim={<left> <lower> <right> <upper>}
 \includegraphics[width=100mm,trim=134 0 110 0 mm, clip=true]{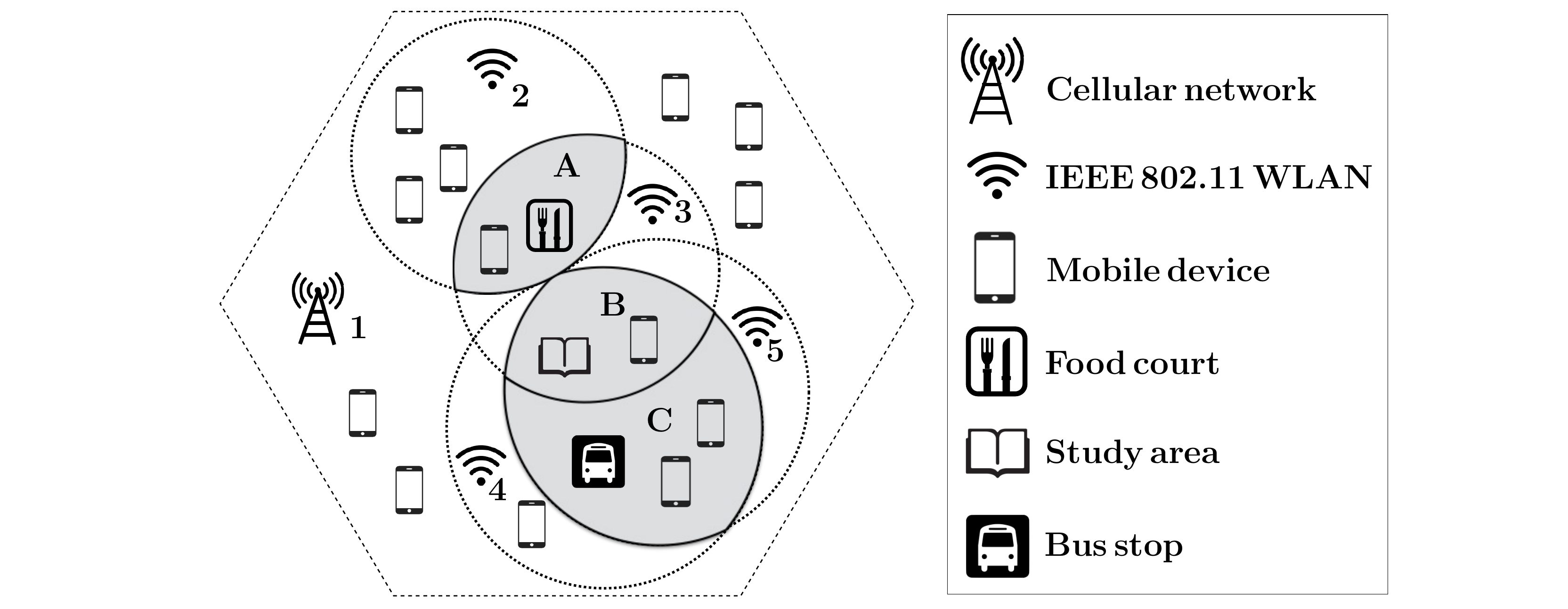}
\end{center}
\caption{Service areas with heterogeneous wireless networks.}
\label{figure:heterogeneousNetworks}
\end{figure}

We consider an environment with multiple wireless devices and heterogeneous wireless networks, such as the one depicted in Figure~\ref{figure:heterogeneousNetworks}. The latter illustrates three service areas, namely a food court, a study area and a bus stop (shaded regions labelled A, B and C, respectively). It shows five wireless networks, numbered 1 to 5, whose area of coverage is delimited by dotted lines. Mobile devices have access to different sets of wireless networks depending on their location. The aim of each device is to quickly identify and associate with the \enquote{optimal} wireless network, which may vary over time. 

To perform an optimal network selection, a device must be aware of the bit rate it can achieve from each network at that time. This is affected by a number of factors, namely (a) the total bandwidth of each network, (b) the distance between the device and the access points (APs) and interference in the environment, and (c) the number of devices associated with each network. A device can learn about a network's bit rate by (a) exploring the network, or (b) estimating it based on feedback received from neighbors who have explored the network. Each time a device switches network, it incurs a cost, which we assume is measured in terms of delay. However, we do not particularly focus on minimizing switching cost in this work; although once the algorithm stabilizes, the frequency of switching networks is expected to be low.

\subsection{Formulation of wireless network selection game}
Mobile devices operate in a dynamic environment, and hence require continuous exploration and adaptation of strategy. Therefore, we formulate the wireless network selection problem as a repeated resource selection game, a special type of congestion game \cite{rosenthal1973class}. We assume that time is slotted and a network selection is made at every time slot.

Since the quality of a network degrades proportionally to its number of associated clients, other mobile devices accessing shared networks can be regarded as adversaries. We consider a setting where devices occasionally share their observations and forward feedback received from others. As such, observations may be received with a delay. We assume that devices are honest and do no lie about their observations. Given the need for sequential decision making, it seems natural to model the behavior of devices using online learning with partial information and delayed feedback in an adversarial setting.

We formally define the wireless network selection game as a tuple 
$\Gamma = \langle \mathcal{N}, \mathcal{K}, (\mathcal{S}_j)_{j \in \mathcal{N}}, (\mathcal{U}^t_i)_{i \in \mathcal{K}}\rangle$, where
\begin{enumerate}
\item $\mathcal{N} = \{1 \cdots n\}$ is the finite set of \textit{n} active (honest) mobile devices indexed by \textit{j}.
\item $\mathcal{K} = \{1 \cdots k\}$ denotes the finite set of \textit{k} wireless networks available 
in the service area. 
\item $\mathcal{S}_j \subseteq 2^{\mathcal{K}_j}$ is the strategy set of mobile device $j$, where $\mathcal{K}_j \subseteq \mathcal{K}$ is the set of networks available to $j$.
\item Gain (payoff or utility) $g_{i,j}(t)$ of device $j$ from network $i$ at time $t$, scaled to [0, 1], is expressed by a function $\mathcal{U}_{i}$ of the number of devices $n_{i}(t)$ associated with $i$ as follows:
$$n_{i}(t) = |\{j' \in \mathcal{N}:i_{j'}(t) = i\}| $$
where $i_{j'}(t)$ is the network selected by $j'$ at time $t$.
%$$ g_{i,j}(t) = \mathcal{U}_{i}^t(n_{i}(t)) $$ 
\begin{align}
 g_{i,j}(t) =
\begin{cases}
\mathcal{U}_{i}^t\left(n_{i}(t)\right) \text{, if } i = i_j(t) \\
\mathcal{U}_{i}^t\left(n_{i}(t) + 1\right) \text{, otherwise.}
\end{cases} \nonumber
\end{align}
If network $i$ has not been explored by device $j$, $g_{i,j}(t)$ may be estimated from gains shared by other devices. 
A device's gain affects its strategy and, hence, ignores switching cost so that networks with high gain but high switching cost are not penalized.
% $g_{i_j, j}(t)$ is the scaled bit rate observed by $j$ from $i_j(t)$. 
% For any $i \in \mathcal{K} - \{i_j(t)\}$ that have been heard of, 
%If network $i$ has not been explored by device $j$, $g_{i,j}(t)$ may be estimated from gains shared by other devices. 
% the reported bit rates from $i$ during time slot $t$.
% 
\item The perceived loss $l_{i,j}(t-\tilde{t},t)$ of network $i$ for device $j$ at time $t - \tilde{t}$ is the difference between the highest bit rate available for $j$ at $t - \tilde{t}$ and the bit rate network $i$ offered, based on what is known by time $t$:
% $$l_{i,j}(t-\tilde{t},t) = \underset{m \in \mathcal{K}}{\max} \{g_{m,j}(t-\tilde{t})\} - g_{i,j}(t-\tilde{t})$$
\begin{align}
 l_{i,j}&(t-\tilde{t},t) \nonumber \\
 &=
\begin{cases}
0 \text{, if }  g_{i,j}(t-\tilde{t}) \text{ is unknown at time $t$,}\\
\underset{m \in \mathcal{K}}{\max} \{g_{m,j}(t-\tilde{t})\} - g_{i,j}(t-\tilde{t}) 
\text{, otherwise.}
\end{cases} \nonumber
\end{align}
% \color{red}in many references, it's computed as 1 - gain\color{black}
% 
\item Cumulative download of a device $j$ is given by $$\sum_{t=1}^{T} \mathcal{U}_{i_j(t)}^t(n_{i_j}(t)) \cdot (slot\_duration - switch\_delay)$$
where $switch\_delay$ is the switching cost ($switch\_delay$ is zero when the device stays in the same network), $slot\_duration$ (higher than $switch\_delay$) is the length of a time slot, and $T$ is the time horizon. For simplicity, we ignore overhead data (packet header) and retransmissions.
\item A strategy profile is given by $\mathcal{S} = \mathcal{S}_1$ x $\cdots$ x $\mathcal{S}_n$. It is at Nash equilibrium \cite{nisan2007algorithmic} if $g_{i_j}(\mathcal{S}) \ge g_{i_j}(\mathcal{S}_{-j}, \mathcal{S}_j')$ for every $\mathcal{S}_j'$ and every $j \in \mathcal{N}$, where $(\mathcal{S}_{-j}, \mathcal{S}_j')$ implies that only device $j$ changes its strategy. Hence, no device wants to unilaterally change its strategy.
\end{enumerate}

The goal of each device is to maximize its cumulative download over time. From a global perspective, we want each mobile device to identify (as quickly as possible) and select the the right network with sufficiently high probability and spend most of the time in it. In other words, we want the algorithm to spend the maximum amount of time at Nash equilibrium.

\section{Co-Bandit} \label{section:proposedAlgorithm}          In this section, we develop Co-Bandit, a novel cooperative bandit algorithm that approximates the EWA (Exponentially Weighted Average) algorithm without communicating full information.
% a cooperative version of the original EWA (Exponentially Weighted Average) algorithm \cite{gyorgy2006adaptive}. 
Co-Bandit is a distributed algorithm and each device $j \in \mathcal{N}$ runs an instance of it. However, the strategy of a device affects those of other devices that have a common set of available networks by affecting their gains. In addition, devices occasionally share their observations to help their neighbors learn faster; instead of each device exploring every network, they cooperatively explore them and share their observations.

We briefly explain EWA, see, e.g. \cite{cesa2006prediction}, and EXP3 \cite{auer2002nonstochastic}. EWA assumes the availability of global knowledge (full information), i.e., a device receives feedback about all networks. On the other hand, EXP3 assumes a bandit setting where a device only receives feedback about its chosen network.

\noindent\textbf{EWA.} 
It maintains a weight for each network, which represents the confidence that the network is a good choice. It starts by assuming uniform weight over all networks. A network's weight is affected by its loss; a lower loss yields a higher weight. Hence, the \enquote{best} network will eventually have highest weight. EWA assumes that time is slotted. At the beginning of every time slot, a device randomly selects a network to associate with during the whole time slot, from a probability distribution which is based on the weights. It observes a bit rate (gain) from the chosen network during the time slot. At the end of the slot, it receives feedback about the gain it could obtain from all other networks. It computes the loss of all networks, updates their weights using a multiplicative update rule. As such, it improves its selection over time.

\noindent\textbf{EXP3.} Much like EWA, it assigns a weight to each network, and initially assumes uniform weight over all networks. A network's weight is affected by its gain; a higher gain yields a higher weight. Therefore, the \enquote{best} network will eventually have highest weight. It assumes that time is slotted. In each time slot, it selects a network at random based on a probability distribution that mixes between using the weights and a uniform distribution; the latter
ensures that EXP3 keeps exploring occasionally and discovers a better network that was previously \enquote{bad}. The device observes a gain (bit rate) from its chosen network. At the end of the time slot, it updates the weight of the chosen network using a multiplicative update rule. 

\noindent\textbf{Difference of Co-Bandit compared to EWA and EXP3.}
% In a full information setting, a device receives feedback about all networks after its selection. On the other hand, in a bandit setting, a device only receives feedback about its chosen network.
Co-Bandit differs from EWA and EXP3 by allowing devices to share their observations with their neighbors. Hence, a device may receive feedback about more than one network, but not necessarily all of them. In addition, it handles feedback received with a delay. We consider a spectrum of settings that lie in between the full information and bandit settings. EWA and EXP3 are applied at the two extremes of the spectrum. As the amount of cooperation increases, the performance of Co-Bandit is expected to be close to that of EWA.
% ; we assume that each device can accurately estimate its gain from networks other than the one it selected based on feedback from neighbors.

We now describe the Co-Bandit algorithm.

\noindent\textbf{Algorithm description.}
Algorithm \ref{algorithm:collaborativeEWA} outlines the major steps in Co-Bandit, excluding the part on 
% reset and 
how to handle a change in the set of available networks. See Table \ref{table:notation} for notations. 
\renewcommand\thempfootnote{\arabic{mpfootnote}}
\begin{table}[!htb]
\small
\centering
\caption{Notations used to describe Co-Bandit
\newline \textit{(subscript $j$ implies that it is specific to a device $j$; $t$ refers to the current time slot, and, when present, indicates that the value is relevant for time slot $t$)}}
\label{table:notation}
\newcolumntype{L}{>{\arraybackslash}m{5.5cm}}
\begin{tabular}{c L}
\hline
$\mathcal{K}_j$             & Set of networks available to $j$.                    \\ 
$k_j$                       & No. of networks available to $j$, i.e., $|\mathcal{K}_j|$.                    \\ 
$n$                         & No. of active mobile devices, i.e., $|\mathcal{N}|$.                    \\ 
$\eta$                      & Learning rate.                            \\
% $\gamma$                    & Frequency to explore \enquote{bad} networks. \\
$w_{i,j}(t)$                & Confidence that network $i$ is a good choice. \\ 
$p_{i,j}(t)$                & Probability for choosing network $i$.         \\ 
$i_j(t)$                    & Network chosen for time slot $t$.             \\ 
$g_{i, j}(t) \in [0,l_{i}]$ & Gain from network $i$.              \\ 
$l_{i,j}(t-\tilde{t},t)$    & Current perceived loss of $j$ from $i$ at $t-\tilde{t}$.\\
$\widehat{l_{i,j}}(t)$      & Loss estimate of network $i$. \\
$\mathbb{I}_{i,j}(t-\tilde{t},t)$ & Indicator function that $i$ was chosen at $t-\tilde{t}$. \\
$q_{i,j}(t-\tilde{t},t)$    & Probability that $\mathbb{I}_{i,j}(t-\tilde{t},t) = 1$.\\
$d$                         & Observations up to $d$ slots old are valid. \\
$p_t$                       & Probability of sharing. \\
$p_l$                       & Probability of listening for messages. \\
$\mathcal{H}_j(t-\tilde{t},t)$        & Devices whose gain for $t-\tilde{t}$ are known.\\
$x$                         & No. of slots a network can be unheard of.         \\ 
$unheard_j$                   & Networks unheard of since time slot $t - x$.      \\ 
\hline
\end{tabular}
\end{table}
\begin{algorithm}[!htb]
  \DontPrintSemicolon % so semicolon is not appended at end of \lif
  \Input{$k_j \in \mathbb{Z}_{>0}$, real $\eta > 0$, $p_t\in [0, 1]$, $p_l\in [0, 1]$, $d \in \mathbb{Z}_{\ge0}$, $x \in \mathbb{Z}_{\ge0}$}
  \Initialize{$ w_{i,j}(1) \leftarrow 1 $ for $ i = 1, \cdots, k_j $ \\ $unheard_j \leftarrow \O$}
  \BlankLine 
  \ForEach{time slot $t = 1, 2, \cdots $}{
    % $ p_{i,j}(t) \leftarrow (1 - \gamma)\frac{w_{i,j}(t)}{\sum\limits_{m=1}^{k} w_{m,j}(t)} + \frac{\gamma}{k} $ for $i = 1, \cdots , k$ \\
    $ p_{i,j}(t) \leftarrow \frac{w_{i,j}(t)}{\sum\limits_{m=1}^{k_j} w_{m,j}(t)}$ for $i = 1, \cdots , k_j$ \\
    \BlankLine
    \If{$unheard_j \neq \O$ and $explore\_unheard()$}{$i_{j}(t) \leftarrow$ random (uniform) from $unheard_j$}
    \lElse{$i_{j}(t) \leftarrow$ random from distribution $p_{j}(t)$}
    \BlankLine
    \Comment{associate with network $i_{j}(t)$}
    \BlankLine
    $g_{i_{j}}(t) \leftarrow $ gain observed, where $g_{i_{j}}(t) \in [0, 1]$\\
    \BlankLine
    \Comment{with probability $p_t$, broadcast observations and messages received}
    \Comment{else, listen with probability $p_l$}
    \Comment{update set $unheard_j$}
    \BlankLine
    \ForEach{ network $i = 1, \cdots, k_j$}{
    $\widehat{l_{i,j}}(t) \leftarrow \dfrac{1}{d'+1}\ \sum\limits_{\tilde{t}=0}^{d'} \dfrac{l_{i,j}(t-\tilde{t}, t)}{q_{i,j}(t-\tilde{t},t)}\ I_{i,j}(t-\tilde{t},t)$
    $$\text{where } d' \leftarrow min\{d,t-1\} \text{; }$$
    % 
    % $$l_{i,j}(t-\tilde{t},t) \leftarrow \underset{m \in \mathcal{K}}{\max} \{g_{m,j}(t-\tilde{t},t)\} - g_{i,j}(t-\tilde{t},t) \text{; }$$
    % 
    % $$g_{i,j}(t-\tilde{t},t) \leftarrow 
    % \begin{cases}
    % g_{i,j}(t-\tilde{t}) \text{, if $i = i_{j}(t-\tilde{t})$},\\
    % mean\ heard \text{, if $i$ is heard of},\\
    % \color{blue}\underset{m \in \mathcal{K}}{\max} \{g_{m,j}(t-\tilde{t},t)\} \text{, otherwise};
    % \end{cases}\color{black}$$
    % 
    $$I_{i,j}(t-\tilde{t},t) \leftarrow \mathbb{I}\{\exists j' \in \mathcal{H}_j(t-\tilde{t},t): i_{j'}(t-\tilde{t}) = i\} \text{; }$$
    $$q_{i,j}(t-\tilde{t},t) \leftarrow 1 - \prod_{j' \in \mathcal{H}_j(t-\tilde{t},t)} (1 - p_{i,j'}(t-\tilde{t}))$$
    \BlankLine
    $ w_{i,j}(t + 1) \leftarrow \dfrac{w_{i,j}(t)\ exp(-\eta\ \widehat{l_{i,j}}(t))}{\underset{m \in \mathcal{K}_j}{\max}\{w_{m,j}(t)\ exp(-\eta\ \widehat{l_{m,j}}(t))\}}$
    }
  }
  \caption{Co-Bandit \BlankLine
  \textit{explore\_unheard() determines whether a device must explore a network unheard of for more than $x$ time slots; it returns True with probability $\frac{|unheard_j|}{n}$; a device explores a network unheard of with probability $\frac{1}{n}$}
  }
  \label{algorithm:collaborativeEWA}
\end{algorithm}

Co-Bandit assumes that time is slotted. Much like EWA and EXP3, it maintains a weight for each network (initially uniform over all networks). A network's weight is affected by its loss, i.e., difference between the highest gain (bit rate) the device could observe during that particular time slot and the bit rate the network had to offer. At the beginning of every time slot, if a device has not learned about some network for a long time, it explores it with some probability. Otherwise, it randomly selects a network from a probability distribution which is based on the weights. It associates with the chosen network for the whole time slot from which it observes some gain. It may decide to share its observation with its neighbors and listen to broadcast messages. As such, it may receive feedback about multiple networks. At the end of the time slot, it updates the weights of all networks; weights of those whose quality are unknown remain unchanged. The same multiplicative weight update and probability update rules as for EWA are used, see, e.g., \cite{cesa2006prediction}, while the loss estimate rule is an adaptation from \cite{cesa2016delay}.

We now further explain the novel aspects of Co-Bandit.

\noindent\textbf{Cooperation.} 
In contrast to a bandit setting, where devices make decisions based solely on their own observations, we consider a setting in which devices cooperate. 
In every time slot, a device observes a gain (bit rate) from its network. It broadcasts its observation with probability $p_t$. Otherwise, it listens for broadcasts with a probability $p_l$. Hence, the underlying communication network is a random directed graph, where the set of vertices denote the mobile devices. A directed edge from device $j$ to device $j'$ implies that $j$ broadcasts its observation and $j'$ listens to and receives the broadcast message. The topology of the graph differs across time slots and depends on the random decisions taken by the devices at different times. Furthermore, it is not known to devices. Cooperation enables devices to leverage feedback received from neighbors to enhance their decisions and speed up their learning rate.

A device's broadcast message includes (a) a timestamp, (b) the device's ID, (c) the network selected, (d) the bit rate observed, (e) an estimate of the number of devices associated with the network selected, (f), the set of networks available, and (g) the device's probability distribution. 
% The format of a broadcast message from device $j$ with its observation at time $t$ is as follows:
% $\langle t, j, i_j(t), \tilde{g}_{i_j}(t), n_{i_j}(t), \mathcal{K}_j,p_j(t) \rangle$.
% 
The timestamp and device ID are used to filter out duplicate messages as they are forwarded (as described next). A device computes its bit rate from a network at time $t$, which it did not explore at that time, based on the bit rate(s) and number of clients of the network reported by neighbors. The probability distribution is used to compute the loss estimate. The set of available networks is useful when devices observe common networks but not necessarily the same set of networks; devices can relate the probability distribution to a set of networks. 

\noindent\textbf{Delayed feedback.} 
Devices not only share their current observation, but everything they have learned during the last $d + 1$ time slots. This includes their own observations as well as feedback received from neighbors. Forwarding messages ensures that it reaches more devices; a device might have missed it earlier as it was not listening, because of packet loss or due to its physical distance from the sender. It also makes it possible for messages to reach many devices while assuming 
% a small value of $p_t$, i.e., transmission probability.
small transmission and listening probabilities.
The order in which messages are received varies and depends on the topology of the random communication graph over the last $d + 1$ time slots. An observation made during time slot $t-d$ can be received any time between $t-d$ and $t$. Hence, the estimate of the loss of a network at time $t - d$ can vary over the last $d + 1$ time slots as new feedback arrives. Observations more than $d$ time slots old are considered stale and are dropped.

\noindent\textbf{Explicit exploration.}
Given that Co-Bandit starts by assuming uniform weight over all networks, all devices may end up perceiving a network with significantly low bandwidth (relative to other networks) as being \enquote{bad}. This may result in no one selecting it. In addition, the quality of a network that was initially \enquote{bad} may improve over time. However, at that time the probability for a device to select it might be too small. To cater for these cases, devices constantly keep track of whether they are learning about all the networks they have access to. They explore those unheard of for a considerable amount of time with probability $\frac{1}{n}$; we do not want all devices to explore it; if they all do, they will most likely observe a low gain. The time must be long enough such that if a device is associated to a network, devices will learn about it. The device exploring a network unheard of broadcasts its observation with probability $1$ so that everyone learns about the network. This feature of the algorithm also implies an additional cost in a setting where a network is actually too \enquote{bad} for anyone to select it.

% \noindent\textbf{Minimal reset.}
% For faster adaptation, the algorithm performs a minimal reset when some specific conditions are met. First, when a device explores an unheard of network and finds it to be better than the one it is selecting with sufficiently high probability (we assume 0.75), Co-Bandit resets the weight of the network being explored (sets it to $1$). The weights of the other networks remain unchanged as the device should not unlearn everything. The aim is to allow a device to identify a better network and quickly start to select it with sufficiently high probability. Second, when a device constantly learns from its neighbors that another network is better than the one it is selecting with sufficiently high probability, it resets the weight of the other network. This allows a device to adapt faster to changes in network quality. It is particularly useful when the difference in bit rates of the two networks is not huge. When resetting, the algorithm also drops data previously received for the last $d$ time slots.

\noindent\textbf{Change in set of networks.}
When a device discovers a new network, e.g., while the mobile user moves around, its weight is set to $1$, i.e., maximum. This will ensure that the device will most likely explore the network. Furthermore, when a network with sufficiently high probability of being selected is no longer available, the weights of all networks are reset. These allow the algorithm to quickly adapt to changes.

\section{Theoretical analysis of Co-Bandit} \label{section:theoreticalAnalysis}
Here, we give an upper bound on the regret of a device using Co-Bandit, and show that Co-Bandit retains the convergence property of multiplicative weight update algorithms with full information. 
%Formal proofs are omitted due to space limitation and provided in the full version of this paper \cite{cobandit2019}.

For the purpose of the analysis, we 
% consider Cooperative EWA without reset and 
assume that (a)
$\mathcal{K}_j = \mathcal{K}$
for every $j \in \mathcal{N}$, i.e., all devices have the same set of networks available to them, (b) the environment is static, and (c) all devices can hear each other, i.e., all devices listening at a particular time will hear those broadcasting at that time.

\noindent\textbf{Regret bound.}
Weak regret refers to the difference between the loss incurred by always selecting the best network in hindsight and that of Co-Bandit. We follow the proof for upper bound on regret for EXP3 \cite{auer2002nonstochastic} and proofs given in \cite{cesa2016delay,alon2017nonstochastic}, and show that Co-Bandit is regret-minimizing.
% has a logarithmic weak regret.

Let $L_{Co\\-Bandit}(T)$ denote the cumulative loss of Co-Bandit at T, $L_{min}(T)$ be the cumulative loss at T when always choosing the best network in hindsight, $d$ be the maximum delay with which a feedback may be received, and $b_0$ be the probability of a device directly hearing from another one.

\begin{theorem}
\label{theorem:regretBound}
For any $k > 0$, 
any maximum delay with which a feedback can be received $d \in \mathbb{Z}_{\ge0}$,
learning rate $\eta =
\sqrt{\frac{max\{\frac{1}{k}, b_0\} \ln k}{e^2 (d + 1) T}}$, 
number of time slots a network can be unheard of $x = d$,
probability of directly hearing from a neighbor $b_0 \in [0, 1 - e^{-1}]$,
% any $\eta \le \frac{1}{ke(d+1)}$,
stopping time $T > (d+1)k\ln k$, and any assignment of rewards, the expected weak regret is upper bounded as:
\begin{align}
\mathbb{E}&\left[L_{Co-Bandit}(T) - L_{min}(T)\right] \nonumber \\
&\hspace{15 mm}\le 2 e \sqrt{\frac{\left(d + 1\right) \ln K \ T}{\max\{\frac{1}{k}, b_0\}}} + d \nonumber
\end{align}
\end{theorem}

\noindent Hence, Co-Bandit is regret-minimizing as its weak regret tends to zero. As time elapses, it performs nearly as well as always selecting the best network in hindsight.

Suppose that we assume $d = 0$, i.e., feedback are received without any delay. When $b_0 = 0$, i.e., no one ever shares its observation, we have a weak regret of order $\sqrt{k \ln k \ T}$ as in the bandit setting \cite{auer2002nonstochastic}. On the other hand, when $b_0 = 1$, i.e. each device always shares its observation, we have a weak regret of order $\sqrt{\ln k \ T}$ as in full information setting, see, e.g. \cite{cesa2006prediction}. 
% \cite{gyorgy2006adaptive}. 
Hence, we can interpolate between full information and bandit settings, special cases of the spectrum of settings considered. 

Assuming that $d > 0$, the higher the delay, the later in time devices may learn about a network's current quality. Hence, the higher the regret bound. However, although not reflected in the regret bound, the higher the value of $d$, the higher the probability of receiving a feedback being forwarded. 
Thus, the better the performance of Co-Bandit in practice.
% Thus, the lower the regret bound.

The formal proof is provided in appendix \ref{appendix:regretBound}

\noindent\textbf{Convergence.}
Strategies in the support of the mixed strategy $\delta_j$ of player $j$ are those played with a non-zero probability \cite{nisan2007algorithmic}. Weakly stable equilibria \cite{kleinberg2009multiplicative} is defined as mixed Nash equilibria $(\delta_1, \cdots, \delta_n)$ with the additional property that each player $j$ remains indifferent between the strategies in the support of $\delta_j$ when any other single player $j'$ changes to a pure strategy in the support of $\delta_{j'}$; however, each strategy in the support of $\delta_j$ may not remain a best response and device $j$ may prefer a strategy outside the support of $\delta_j$.

We prove that Co-Bandit retains the convergence property of multiplicative-weights learning algorithm in a full information setting \cite{kleinberg2009multiplicative} and EXP3 \cite{tekin2011performance}. 

\begin{theorem}
When $\eta$ is arbitrarily small and 
% the frequency of exploring some network unheard of is arbitrarily small, 
the number of devices $n$ tends to infinity (frequency of exploring network(s) unheard of tends to zero), 
the strategy profile of all devices using Co-Bandit converges to a weakly stable equilibrium; weakly stable equilibria are pure Nash equilibria with probability $1$ when the bit rate of each network is chosen at random independently \cite{kleinberg2009multiplicative}.
\end{theorem}

Hence, when all devices leverage Co-Bandit, they end up being optimally distributed across networks. No one will observe higher gain by unilaterally switching network. We show that the dynamics of the probability distribution over the set of networks is given by the following replicator equation:
\begin{align}
\xi_{i,j} 
&= p_{i,j} \sum\limits_{m \in \mathcal{K} - \{i\}} p_{m,j} \left(l_{m,j} - q\ l_{i,j}\right) \nonumber
\end{align}

In the full information setting, the probability $q$ of hearing about network $i$ is equal to $1$. In that case, we get a replicator dynamic identical to the one in \cite{kleinberg2009multiplicative}. With a drop in the value of $q$, the value of $\xi_{i,j}$ will rise, implying a slower convergence. Furthermore, in the bandit setting, $b = 0$ and $q = p_{i,j}$. Given our definition of loss, $\xi_{i,j}$ will be zero all the time. Hence, the algorithm never converges to the right state. As such, we interpolate between the full information and bandit settings.

The formal proof is provided in appendix \ref{appendix:convergence} 

\section{Implementation details} \label{section:implementation}
    We evaluate Co-Bandit and compare its performance against those of EWA, see, e.g. \cite{cesa2006prediction}, and EXP3 \cite{auer2002nonstochastic}, through simulation using synthetic data. 
% They are  algorithms for the two extremes of the spectrum of settings considered.
% Table \ref{table:algorithm_comparison} briefly outlines how EXP3 and Smart EXP3 work. 
All algorithms are implemented in Python, using SimPy \cite{simpy}. We discuss the implementation of Co-Bandit and specify parameter values chosen. 

% \input{table/algorithm_comparison}
% focusing on the parameter values chosen and minimal reset. 

% \noindent\textbf{Parameter choice.} 
\noindent\textbf{Learning rate.} 
In our implementation of Co-Bandit, learning rate $\eta = 10$. In the theoretical analysis, we assume a small value for $\eta$. However, in practice, we observe that Co-Bandit takes too long to stabilize at the optimal state when $\eta$ is very small; but, assuming a very high learning rate makes Co-Bandit too \enquote{aggressive} and it stabilizes at a sub-optimal state. 

\noindent\textbf{Cooperation.}
Unless specified otherwise, each device shares its knowledge (transmits) with probability $p_t = \frac{1}{n}$. Otherwise, it listens for broadcast messages with probability $p_l = \frac{1}{3}$, i.e. a device listens for feedback messages once every three time slots. We assume here that devices can estimate the number of devices associated to its network, e.g., from feedback received over time or scanning for arp messages \cite{arp_scan} for WiFi.

\noindent\textbf{Delayed message.}
We assume $d = 5$, i.e., messages up to $5$ time slots old are considered valid. While increasing the value of $d$ raises the likelihood of devices receiving feedback, it also implies more data has to be stored and broadcast. In addition, in a dynamic wireless network setting, old observations may be more misleading than useful. Devices drop duplicate messages received as they are forwarded.

\noindent\textbf{Gain.} 
Although it is not a pre-requirement for Co-Bandit, for simplicity, we assume that a network's bandwidth is equally shared among its clients. While the gain of $i_j(t)$ is the scaled bit rate device $j$ observes from network $i$ at time $t$, the gain of a network $i \in \mathcal{K}_j - \{i_j\}$ is estimated from feedback $j$ receives from its neighbors. We compute the unscaled gain of network $i$ as 
% $g_{i,j}(t) = 
$\frac{1}{n_i(t) + 1}\ \sum_{j' \in \mathcal{H}_j: i_{j'}(t) = i} \ g_{i, j'}(t)$.

\noindent\textbf{Switching cost.}
We model delay (switching cost) using Johnson's SU distribution for WiFi and Student's t-distribution for cellular, each identified as a best fit~\cite{fitter} to 500 delay values collected from real world experiments.

\noindent\textbf{Exploration.}
Devices explore networks unheard of for 32 time slots or more (i.e., $x = 32$, or 8 simulated minutes) with probability $\frac{1}{n}$; we assume $n$ can be estimated. While we assume, $x = d$ in the theoretical analysis, in practice we assume a higher value for $x$, as we want to limit the frequency of exploring networks which may cause the algorithm to spend most of the time in a sub-optimal state. In addition, $x \ge k_j$ since we hear from one devices (on average) in every time slot. However, if $x$ is too big, the rate of adaptation drops. 
% When there are multiple networks unheard of, the device explores any one of them uniformly at random.

% \noindent\textbf{Minimal reset.}
% When a device explores a network $i'$ and observes a bit rate higher than the median (over the last $x$ time slots) bit rate observed from the network, say $i$, that it selects with probability $0.75$ or more, it resets the weight of $i'$. In addition, after having access to the networks (merely seeing them) for at least $x$ time slots, once every $\frac{x}{2}$ time slots, each device $j$ reviews the recent quality of each network. If any network $i'$ had higher bit rate than network $i$ which $j$ selects with probability at least $0.75$, then $j$ resets the weight of $i'$ with probability $\frac{1}{n_{i}}$; $n_{i}$ is an estimate of network $i$'s number of clients when $j$ last associated with it. Given that the real world is noisy, it might be better to consider resetting a weight only if the network offers a bit rate of a certain percentage higher to prevent frequent resets.

% \noindent\textbf{EXP3 and its parameter choice.} It assigns a weight to each network. In each time slot, it selects a network at random based on a probability distribution that mixes between using the weights and a uniform distribution. At the end of a time slot, the device observes a gain (bit rate) from its chosen network and updates its weight using a multiplicative update rule. 
\noindent\textbf{Parameter choice for EWA and EXP3.} 
For EWA, we assume the same learning rate as for Co-Bandit, i.e., $\eta = 10$. We assume that $\gamma = t^{-\frac{1}{3}}$ for EXP3 \cite{maghsudi2013relay}.
% 
% For Smart EXP3, we assume that (a) $\gamma = t^{-\frac{1}{3}}$ (as for EXP3), (b) $\beta = 0.1$, (c) device $j$ selects greedily with probability $\frac{1}{2}$ when $max(p_j) - min(p_j) \le \frac{1}{k - 1}$ or the block length of the network selected with highest probability is low, 
% (d) it switches to the previous network if the average gain of the last 8 time slots, or gain from last time slot, or more than 50\% of gain from the last 8 time slots is higher than the current gain, and (e) it resets when a device stays for more than $40$ time slots in the network which has a probability of at least $0.75$ or if a drop of at least 15\% is observed in the network being selected with the highest frequency (and since more than 4 time slots).

\section{Evaluation in static settings} \label{section:evaluationStaticSetting}
    In this section, we evaluate Co-Bandit in static settings, i.e. where the number of mobile devices in the service area, and the number and quality of wireless networks remain constant. We compare its performance to those of EWA and EXP3. We show that (a) as the amount of cooperation increases, its performance approaches that of EWA in a full information setting, (b) delayed feedback ensures that information reaches more devices, enhancing performance of the algorithm, (c) it stabilizes at Nash equilibrium relatively fast (comparable to that of EWA), (d) it far outperforms EXP3 in terms of rate of stabilization and per device cumulative download, and (e) it scales with an increase in number of devices and number of wireless networks in the service area. 

In our prior work, we proposed Smart EXP3 \cite{appavoo2018shrewd}, a bandit algorithm with far better practical performance than EXP3. However, we exclude comparison of the performance of Co-Bandit to that of Smart EXP3 as the latter has several features that we can add to Co-Bandit to improve its performance.

\noindent\textbf{Setup.}
We observe very good performance with few networks; it gets more challenging when we increase the number of networks. Thus, we consider settings with 20 devices and 5 networks; in general, most places will not have more networks. We assume non-uniform data rates 18, 8, 13, 16 and 10 Mbps,
% 6, 7, 22, 16 and 14 Mbps, 
a factor close to the theoretical data rates of IEEE 802.11 standards \cite{ieee80211} and cellular networks \cite{gessner2012umts} that yields a unique Nash equilibrium, unless specified otherwise. Although it is not a pre-requirement for the algorithm, we assume that (a) mobile devices are time-synchronized, and (b) all devices can hear each other. Results involve data from %500
100 
runs of 5 (simulated) hours each, i.e., 1200 time slots, unless specified otherwise.

\noindent\textbf{Evaluation criteria.} We evaluate the performance of each algorithm based on (a) the state at which it stabilizes and how \enquote{bad} this is compared to Nash equilibrium; we use the notions of stability and distance to Nash equilibrium from \cite{appavoo2018shrewd} (Definitions 2 and 3), (b) the time it takes to stabilize, (c) cumulative download of the devices, and (d) scalability. 

An algorithm is said to have reached a stable state when each device selects a particular network with probability at least 0.75 until the end (at least over the last 10 time slots). EWA is not inherently stable (as per our definition of stability). For example, when a device observes the same gain from two networks, it selects them with equal probability. 

Distance to Nash equilibrium is the maximum percentage higher gain any device would have observed if the algorithm was at Nash equilibrium, compared to its current gain. If no device can achieve more than $\epsilon$ percent increase in gain by unilaterally deviating from its strategy, then the algorithm is at $\epsilon-equilibrium$ \cite{nisan2007algorithmic}.

% Strategy profile $\mathcal{S} = \mathcal{S}_1$ x $\cdots$ x $\mathcal{S}_n$ is at $\epsilon-equilibrium$ \cite{nisan2007algorithmic} if $g_{i_j}(\mathcal{S}) \ge g_{i_j}(\mathcal{S}_{-j}, \mathcal{S}_j') - \epsilon$ for every $\mathcal{S}_j'$ and every $j \in \mathcal{N}$, where $i_j$ is the network selected by device $j$, $g_{i_j}(\mathcal{S})$ is the gain observed by device $j$ given strategy profile $\mathcal{S}$, $\epsilon$ is a real non-negative parameter and $(\mathcal{S}_{-j}, \mathcal{S}_j')$ implies that only device $j$ changes its strategy. Thus, no device can achieve more than $\epsilon$ increase in gain by unilaterally deviating from its strategy.

\begin{figure}[!htb]
\begin{center}
\includegraphics [scale=1]
{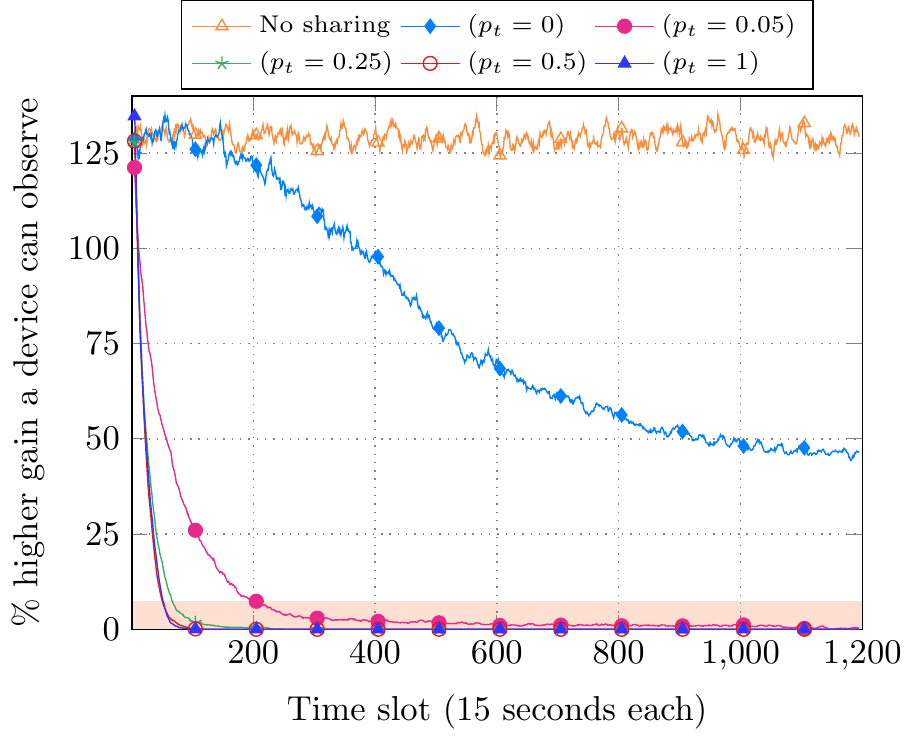}
\end{center}
\caption{Tradeoff between the amount of communication and average distance to Nash equilibrium of Co-Bandit (\% higher gain any device would have observed, compared to its current gain, if the algorithm was at Nash equilibrium) --- shaded region represents $\epsilon$-equilibrium, where $\epsilon = 7.5$.}
\label{figure:cooperation_distanceToNashEquilibrium}
\end{figure}

\noindent\textbf{Effect of cooperation.}
    We study the effect of cooperation by varying the probability of sharing, i.e., the value of $p_t$. Here, we ignore delayed feedback, i.e., $d = 0$, and assume that devices always listen, even while transmitting. Figure \ref{figure:cooperation_distanceToNashEquilibrium} shows that the performance of Co-Bandit improves as the value of $p_t$ increases, as expected (shown by a lower distance to Nash equilibrium). The distance also drops when $p_t = 0$ as devices were still sharing their observation when exploring a network unheard of; yielding a better performance than that of EXP3 (discussed later) even with a small amount of information occasionally. When devices never share anything (\enquote{No sharing}), they only observe a gain from their chosen network. Hence, given our update rules, their probability distribution remains uniform, and the distance never drops.

All the runs stabilized at Nash equilibrium when $p_t \ge 0.05$ (considering the values of $p_t$ in Figure \ref{figure:cooperation_distanceToNashEquilibrium}). When $p_t = 0$, 8 runs were stable at Nash equilibrium, 8 runs did not stabilize, and the other runs stabilized at a state that require a maximum of 2 devices to switch network to reach Nash equilibrium. Table \ref{table:cooperation_stabilization_rate} shows the time Co-Bandit takes to stabilize in terms of median number of time slots, given the amount of cooperation. As expected, the time to stabilize decreases as the amount of cooperation rises (denoted by an increase in the value of $p_t$).

\begin{table}[!htb]
\small
\centering
\caption{Effect of cooperation on the time Co-Bandit takes to stabilize (whether at Nash equilibrium or some other state).}
\label{table:cooperation_stabilization_rate}
\newcolumntype{L}{>{\centering\arraybackslash}p{4.5cm}}
\begin{tabular}{l L}
\hline
\textbf{Amount of cooperation} & \textbf{\# time slots to stabilize (median)} \\ \hline
\multicolumn{1}{l}{\textbf{No sharing}} & - \\ 
\multicolumn{1}{l}{\textbf{\boldmath$p_t = 0$}} & 720.5 \\ 
\multicolumn{1}{l}{\textbf{\boldmath$p_t = 0.05$}} & 143 \\ 
\multicolumn{1}{l}{\textbf{\boldmath$p_t = 0.25$}} & 57 \\ 
\multicolumn{1}{l}{\textbf{\boldmath$p_t = 0.5$}} & 45.5 \\ 
\multicolumn{1}{l}{\textbf{\boldmath$p_t = 1$}} & 48 \\ 
\hline
\end{tabular}
\end{table}

% Figure \ref{figure:cooperation_stable_state} shows a drop in percentage of runs that stabilized at Nash equilibrium (or any state) as the probability of sharing decreases. When $p_t = 1$, 5 (out of 500) runs did not stabilize at Nash equilibrium. In full information setting, global knowledge is assumed. When $p_t = 1$, devices are still unaware of the quality of networks to which no one is associated; we observe that 4 runs stabilized at a state with no one in network 1 (which would have been optimal if network 1 was not available). The other run required a single device to switch network to be at Nash equilibrium; the device would then observe an increase in bit rate of only 0.2 Mbps. 
% Once the algorithm selects a network with sufficiently high probability, without a minimal reset, adaptation is slow when another network is only slightly better.
% \begin{figure}[!htb]
% \begin{center}
% \includegraphics [scale=0.78]
% {plot/cooperation_stable_state.pdf}
% \end{center}
% \caption{Tradeoff between the amount of communication and stability (whether a stable state was reached, and the type of such state, i.e., Nash equilibrium or some other state).}
% \label{figure:cooperation_stable_state}
% \end{figure}
    
\begin{figure}[!htb]
\begin{center}
\includegraphics %[scale=0.84]
{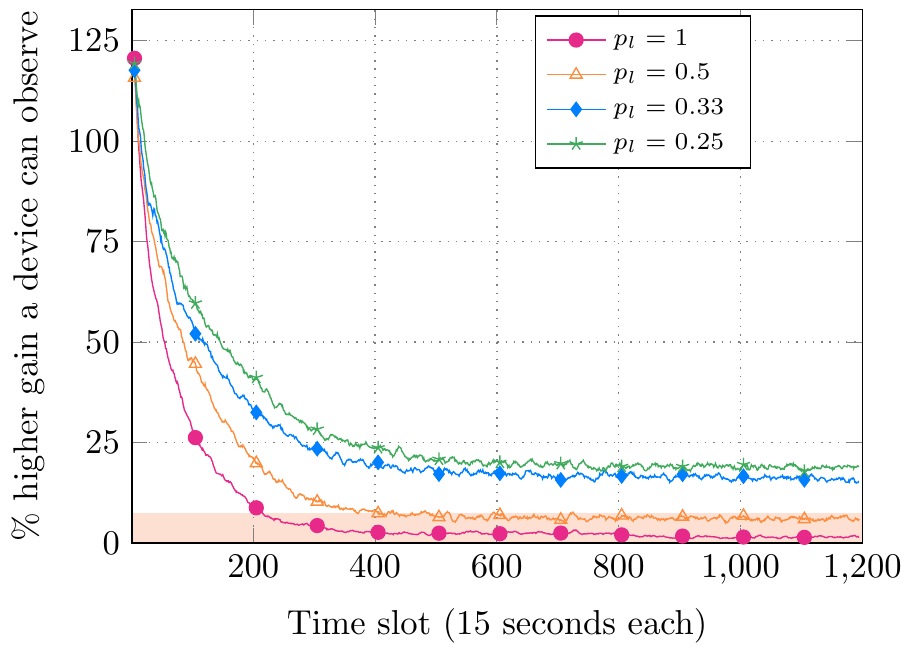}
\end{center}
\caption{Effect of varying probability of listening on average distance to Nash equilibrium of Co-Bandit (\% higher gain any device would have observed, compared to its current gain, if the algorithm was at Nash equilibrium) --- shaded region represents $\epsilon$-equilibrium, where $\epsilon = 7.5$.}
\label{figure:different_listen_probability_distanceToNashEquilibrium}
\end{figure}

\noindent\textbf{Effect of delayed feedback.}
    We want Co-Bandit to quickly stabilize at Nash equilibrium with minimal communication; a device should not spend all the time broadcasting or listening for feedback. We assume $p_t=\frac{1}{n}$, i.e. (on average) one device communicates in every time slot. We ignore delayed feedback, i.e. set $d = 0$, and vary the probability with which a device listens for feedback. If a device is not broadcasting, it listens with probability $p_l$. Figure \ref{figure:different_listen_probability_distanceToNashEquilibrium} shows that as $p_l$ decreases, the performance drops (distance to Nash equilibrium increases).
% \begin{figure}[!htb]
% \begin{center}
% \includegraphics [scale=0.84]
% {plot/different_listen_probability_distanceToNashEquilibrium.pdf}
% \end{center}
% \caption{Effect of varying probability of listening on average distance to Nash equilibrium of Co-Bandit (\% higher gain any device would have observed, compared to its current gain, if the algorithm was at Nash equilibrium) --- shaded region represents $\epsilon$-equilibrium, where $\epsilon = 7.5$.}
% \label{figure:different_listen_probability_distanceToNashEquilibrium}
% \end{figure}

\begin{figure}[!htb]
\begin{center}
\includegraphics %[scale=0.84]
{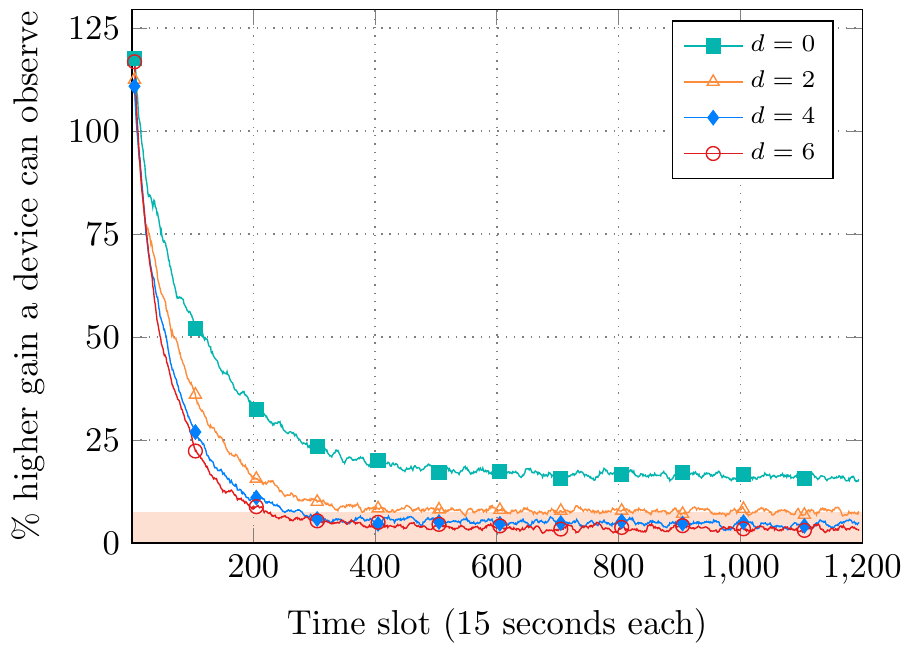}
\end{center}
\caption{Tradeoff between delay in feedback and average distance to Nash equilibrium of Co-Bandit (\% higher gain any device would have observed, compared to its current gain, if the algorithm was at Nash equilibrium) --- shaded region represents $\epsilon$-equilibrium, where $\epsilon = 7.5$.}
\label{figure:delayed_feedback_distanceToNashEquilibrium}
\end{figure}

\begin{figure}[!htb]
\begin{center}
\includegraphics %[scale=0.84]
{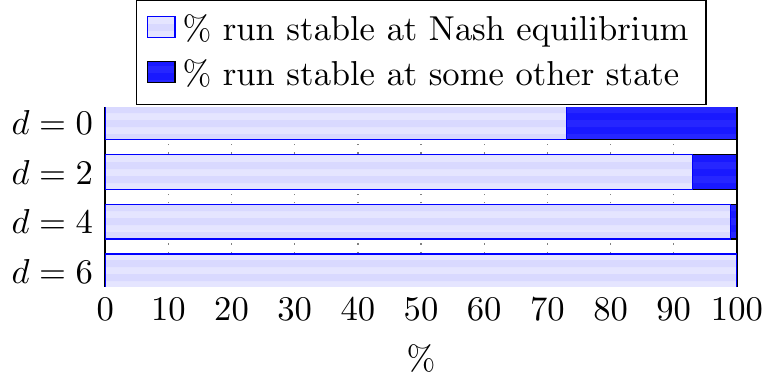}
\end{center}
\caption{Tradeoff between how much delay in feedback is acceptable and stability of Co-Bandit (whether stable and type of stable state --- Nash equilibrium or some other state).}
\label{figure:delayed_feedback_stable_state}
\end{figure}

We leverage delayed feedback, where devices communicate everything they have learned since time slot $t - d$ to ensure that feedback reaches more devices, even if they transmit and listen with a small probability. We assume that devices listen with probability $\frac{1}{3}$ and evaluate the effect of delayed feedback. We observe, from Figures \ref{figure:delayed_feedback_distanceToNashEquilibrium} and \ref{figure:delayed_feedback_stable_state}, that Co-Bandit is stable at a better state (on average) as the value of $d$ increases. Figure \ref{figure:delayed_feedback_distanceToNashEquilibrium} shows that the distance to Nash equilibrium drops as $d$ increases; it also stabilizes faster. 

Figure \ref{figure:delayed_feedback_stable_state} shows that the percentage of runs that are stable at Nash equilibrium rises as $d$ increases. However, $d$ should not be too high as the quality of networks may change quickly rendering observations far back in time irrelevant. When $d>0$, we notice that all runs which are stable at a state other than Nash equilibrium requires a single device to switch network to reach Nash equilibrium.

\begin{figure}[!htb]
\begin{center}
\includegraphics %[scale=0.84]
{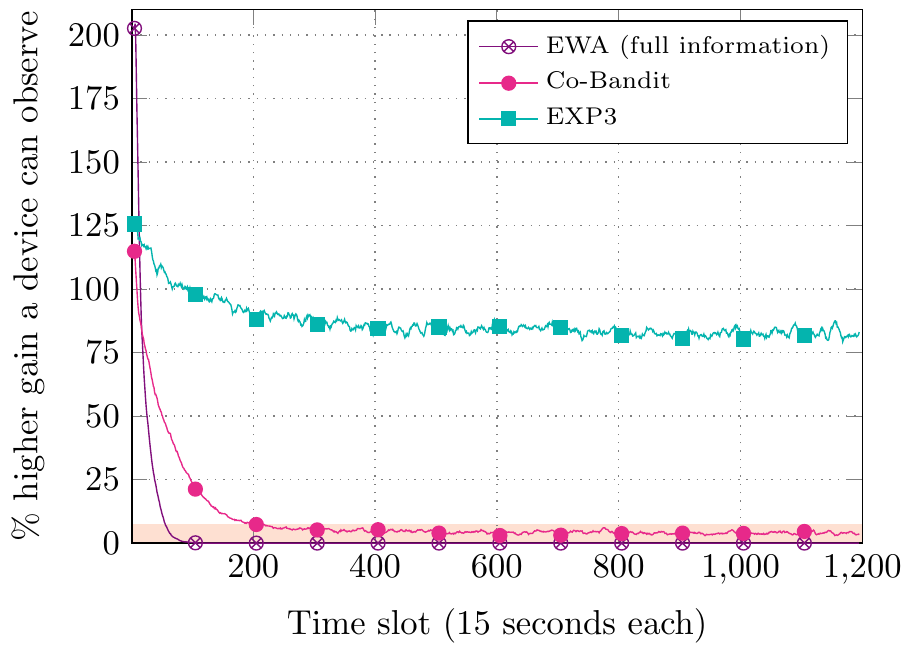}
\end{center}
\caption{Performance comparison of Co-Bandit, EWA and EXP3 based on distance to Nash equilibrium (\% higher gain any device would have observed, compared to its current gain, if the algorithm was at Nash equilibrium)--- shaded region represents $\epsilon$-equilibrium, where $\epsilon = 7.5$.}
\label{figure:algorithm_comparison_distanceToNashEquilibrium}
\end{figure}

\noindent\textbf{Performance comparison of Co-Bandit, EWA, and EXP3.}
    Figure \ref{figure:algorithm_comparison_distanceToNashEquilibrium} shows that Co-Bandit far outperforms EXP3. It always stabilized at Nash equilibrium. Yet, Co-Bandit maintains a small non-zero distance after stabilizing. This is due to the cost of exploring networks unheard of for a significant amount for time. While all runs of EWA were stable at Nash equilibrium, none of those for EXP3 stabilized within 1200 time slots. In some simulations with 20 devices and 3 wireless networks, EXP3 took over 85,500 time slots (on average) to stabilize at Nash equilibrium \cite{appavoo2018shrewd}. Table \ref{table:algorithm_comparison_cumulative_download_stabilization_rate} shows that Co-Bandit achieves a median cumulative gain that is comparable to that of EWA, but 45\% higher than that of EXP3. Co-Bandit took 2.69x more time (median) than EWA to 
stabilize.

\begin{table}[!htb]
\small
\centering
\caption{Median per device cumulative download (GB) and median time Co-Bandit, EWA and EXP3 take to stabilize (whether at Nash equilibrium or some other state).}
\label{table:algorithm_comparison_cumulative_download_stabilization_rate}
\newcolumntype{L}{>{\centering\arraybackslash}p{2.22cm}}
\begin{tabular}{l L L}
\hline
\textbf{Algorithm} & \textbf{Cumulative download (GB)} & \textbf{\# time slots to stabilize} \\ \hline
\multicolumn{1}{l}{\textbf{EWA (full information)}} & 7.02 & 50 \\ 
\multicolumn{1}{l}{\textbf{Co-Bandit}} & 6.96 & 134.5 \\ 
\multicolumn{1}{l}{\textbf{EXP3}} & 4.80       & -       \\ 
% \multicolumn{1}{l}{\textbf{Smart EXP3 w/o reset}} & 6.88 & 754 \\ 
\hline
\end{tabular}
\end{table}

\begin{figure}[!htb]
\begin{center}
\includegraphics %[scale=0.84]
{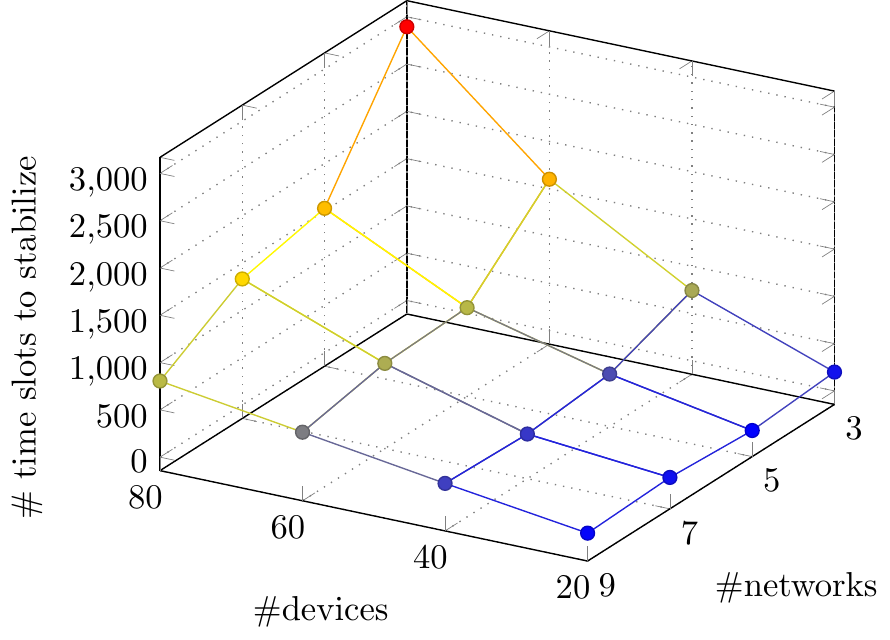}
\end{center}
\caption{Scalability of Co-Bandit with an increase in number of devices and/or networks --- in terms of rate of stabilization.}
\label{figure:scalability}
\end{figure}

\noindent\textbf{Scalability.}
    We evaluate the scalability of Co-Bandit in terms of the rate at which it reaches a
stable state. The algorithm was run 50 times, for 20,000 time slots (i.e., 83.33 simulated hours) each, with different numbers of devices and networks. It always stabilized, nearly all the time at either Nash equilibrium or a state that requires a single device to switch network to be at Nash equilibrium. Figure \ref{figure:scalability} shows that as the number of devices grows, Co-Bandit takes more time to stabilize. The effect is even more significant with 3 networks, as the gain and loss observed are very small when the number of devices is high. With 20 devices, a rise in number of networks does not affect stabilization time much. But, with higher number of devices, a rise in number of networks yields faster stabilization.

\begin{figure}[!htb]
\begin{center}
\includegraphics %[scale=0.84]
{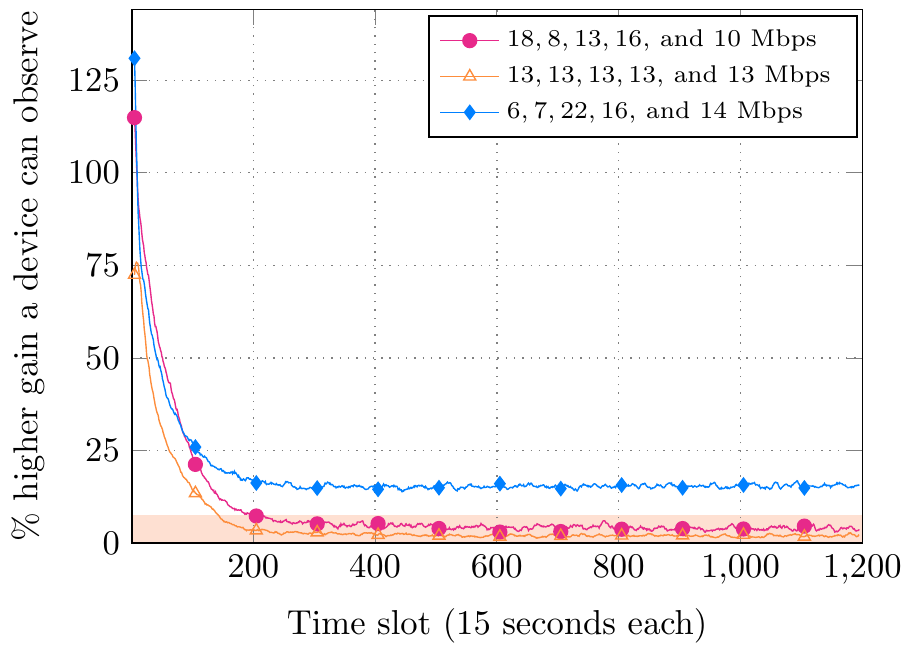}
\end{center}
\caption{Distance to Nash equilibrium of Co-Bandit (\% higher gain any device would have observed, compared to its current gain, if the algorithm was at Nash equilibrium) in setups with 20 devices and 5 networks and aggregate bandwidth 65 Mbps --- shaded region represents $\epsilon$-equilibrium, where $\epsilon = 7.5$.}
\label{figure:different_data_rates_distanceToNashEquilibrium}
\end{figure}

\noindent\textbf{Setups with other data rates.}
    We evaluate Co-Bandit in two additional setups with 20 devices and 5 networks, with an aggregate bandwidth of 65 Mbps that yields a unique Nash equilibrium. Figure \ref{figure:different_data_rates_distanceToNashEquilibrium} shows that, in both setups, Co-Bandit stabilized. 
Table \ref{table:different_data_rates_cumulative_download_stabilization_rate} gives the cumulative download and time taken to stabilize in each setup.
In the first setup, the networks have a uniform data rate of 13 Mbps each. All runs stabilized at Nash equilibrium. Given that the algorithm starts by assuming uniform weight over all networks, it stabilized faster in this setup.
The second setup assumes non-uniform data rates 6, 7, 22, 16 and 14 Mbps. Here, the optimal distribution of devices is more skewed. Hence, Co-Bandit takes longer to stabilize. 44\% run were stable at Nash equilibrium. Most of the other runs were stable at a state with no device in a network or which requires a single device to switch network to observe a slightly higher gain (up to 0.25 Mbps) for Co-bandit to be at Nash equilibrium. 
% setup 1
% 18_8_13_16_10
% 6_2_4_5_3
% setup 2
% 13_13_13_13_13
% 4_4_4_4_4
% setup 3
% 6_7_22_16_14
% 2_2_7_5_4

\begin{table}[!htb]
\small
\centering
\caption{Median per device cumulative download (GB) and median time Co-Bandit takes to stabilize (whether at Nash equilibrium or some other state) in setups with 20 devices and 5 networks with an aggregate bandwidth of 65 Mbps.}
\label{table:different_data_rates_cumulative_download_stabilization_rate}
\newcolumntype{L}{>{\centering\arraybackslash}p{2.22cm}}
\begin{tabular}{l L L L}
\hline
\textbf{Data rates (Mbps)} & \textbf{Cumulative download (GB)} & \textbf{\# time slots to stabilize} \\ \hline
\multicolumn{1}{l}{\textbf{18, 8, 13, 16, 10}} & 6.96 & 134.5 \\ 
\multicolumn{1}{l}{\textbf{13, 13, 13, 13, 13}} & 7.06 & 114.5 \\ 
\multicolumn{1}{l}{\textbf{6, 7, 22, 16, 14}} & 6.87 & 175 \\ 
\hline
\end{tabular}
\end{table}

We now consider a minimal reset to improve the performance of Co-Bandit in setups which require a skewed optimal distribution of devices across networks. 

\noindent\textbf{Effect of minimal reset.}
    % In our prior work, we proposed Smart EXP3 \cite{appavoo2018shrewd}, a major improvement over EXP3. It runs independently on each device, maintains a weight for each network, and uses the same update rules as EXP3. However, it (a) starts by exploring each network in random order, (b) selects a network for a block (consecutive time slots) whose length increases with the frequency of selection of the network, (c) occasionally makes a deterministic (greedy) selection, (d) allows a device to switch back to its previous network upon selecting a worse network, and (e) performs a minimal reset periodically and when a significant drop is detected in the network being selected with sufficiently high probability. 
% 
We evaluate the effect of a minimal reset in the setup with data rates 6, 7, 22, 16 and 14 Mbps. First, when a device explores a network unheard of and finds it to be better than the one it is selecting with probability at least 0.75, Co-Bandit resets the weight of the network being explored. The weights of other networks remain unchanged as the device must not unlearn everything. The aim is to allow a device to identify a better network and quickly adapt. Second, when a device constantly learns from its neighbors that another network is better than the one it is selecting with probability at least 0.75 (by more than a percentage; we assume $2.5\%$), it resets the weight of the other network with probability $\frac{1}{n_{i_j}}$ ($n_{i_j}$ is estimated). This helps when the difference in bit rates of the two networks is small, and allows a device to adapt faster to changes in network quality. When resetting, Co-Bandit drops data previously received for the last $d$ time slots.

\begin{figure}[!htb]
\begin{center}
\includegraphics %[scale=0.84]
{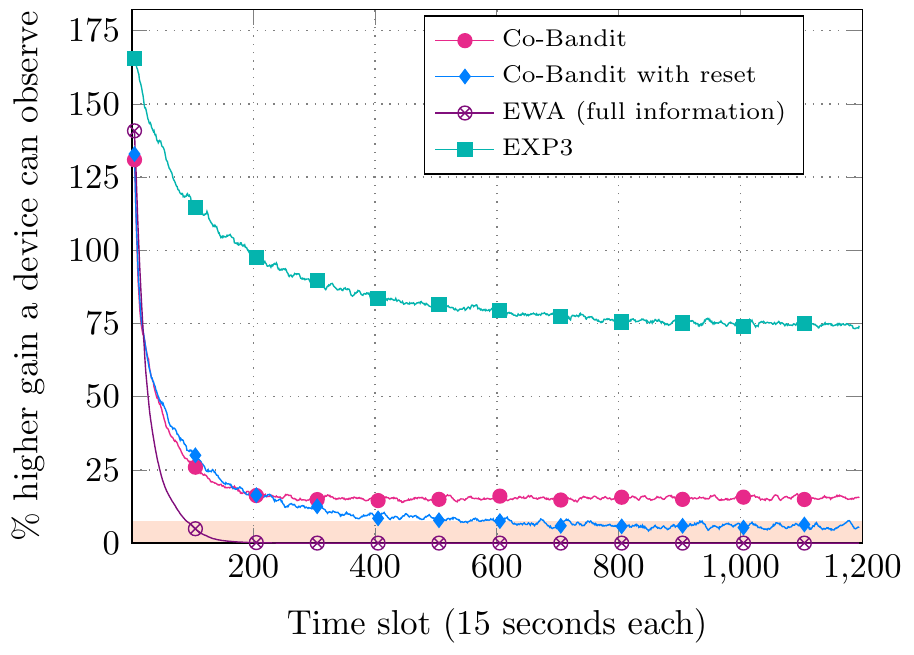}
\end{center}
\caption{Effect of minimal reset on average distance to Nash equilibrium of Co-Bandit (\% higher gain any device would have observed, compared to its current gain, if the algorithm was at Nash equilibrium)--- shaded region represents $\epsilon$-equilibrium, where $\epsilon = 7.5$.}
\label{figure:minimal_reset_distanceToNashEquilibrium}
\end{figure}

Figure \ref{figure:minimal_reset_distanceToNashEquilibrium} shows that a minimal reset improves performance (on average) with a higher percentage of runs (96\%) stable at Nash equilibrium. The other runs did not stabilize. But, it increases stabilization time by approximately 3.4x (median).
%Table \ref{table:minimal_reset_cumulative_download_stabilization_rate} shows that while the minimal reset increases the time Co-Bandit takes to stabilize, it increases the number of runs stable at the optimal state. The cumulative gain of Smart EXP3 is lower due to its periodic reset, denoted by the spikes. For the same reason, the time to stabilize does not apply in this case.

% Table \ref{table:minimal_reset_cumulative_download_stabilization_rate} shows the median cumulative gain of a device and the median time each algorithm takes to stabilize (whether to Nash equilibrium or some other state). The median cumulative gain of a device for Cooperative EWA is approximately same as that of EWA, slightly higher than that of Smart EXP3 without reset, and significantly (about 44\%) higher than that of EXP3. The median time Cooperative EWA takes to stabilize is approximately 1.8 times higher than that of EWA but about 4.5 times less than that of Smart EXP3. The version of the algorithm without delayed feedback and reset takes (median) 198 time slots to stabilize while the version with delayed feedback but without reset takes (median) 144 time slots. While the minimal reset increases the time the algorithm takes to stabilize it ensures that it is stable at the optimal state. 

%\input{table/minimal_reset_cumulative_download_stabilization_rate}

% EXP3, Smart EXP3, EWA data is from 500 runs

% \noindent\textbf{Devices listening with other probabilities.}
    % \input{evaluation/otherListenProbability}

% \noindent\textbf{Multiple selection algorithms.}
%     \input{evaluation/multipleAlgorithms}

\section{Evaluation in dynamic settings} \label{section:evaluationDynamicSetting}
    In this section, we evaluate the adaptability of Co-Bandit to changes in the environment, namely when (a) devices join and leave the service area at different times, and (b) the set of networks available and quality of networks change over time as users of mobile devices move across service areas. We compare its performance to those of EWA and EXP3, and show that it gracefully adapts to changes, with a performance comparable to that of EWA. It far outperforms EXP3.

We consider three dynamic settings involving 20 mobile devices. As evaluation criteria, we consider the states of the algorithms over time and how far these states are from Nash equilibrium. As in Section \ref{section:evaluationStaticSetting}, we assume that mobile devices are time-synchronized (although not a pre-requirement for the algorithm). We present results of Co-Bandit, both with and without the minimal reset. All results involve data from 100 runs of 5 (simulated) hours each, i.e., 1200 time slots.

\begin{figure}[!htb]
\begin{center}
\includegraphics %[scale=0.9]
{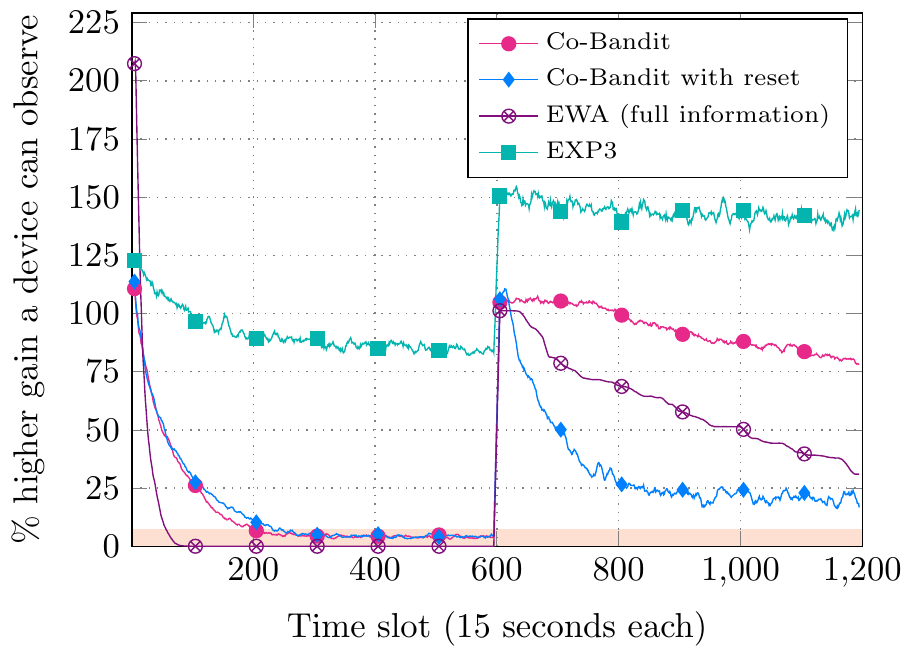}
\end{center}
\caption{Average distance to Nash equilibrium (\% higher gain any device would have observed, compared to its current gain, if the algorithm was at Nash equilibrium) in a setting where 10 devices leave the service area at the end of t = 600 --- shaded region represents $\epsilon$-equilibrium, where $\epsilon = 7.5$.}
\label{figure:dynamic_setting_1_distanceToNashEquilibrium}
\end{figure}

\noindent\textbf{Devices leaving the service area.}
    In this setting, all 20 mobile devices have access to the same 5 wireless networks with data rates 18, 8, 13, 16 and 10 Mbps, as in Section \ref{section:evaluationStaticSetting}. We assume that all devices can hear each other. 10 of the devices leave the service area at the end of time slot t = 600, freeing resources. Figure \ref{figure:dynamic_setting_1_distanceToNashEquilibrium} shows that Co-Bandit dynamically adapts to the change, and far outperforms EXP3. With the minimal reset, Co-Bandit adapts faster than EWA when resources are freed at t = 600. Without a reset, EWA takes time adapt once it is stable at another state (here, the optimal state when 20 devices were in the service area). 

\begin{figure}[!htb]
\begin{center}
\includegraphics %[scale=0.9]
{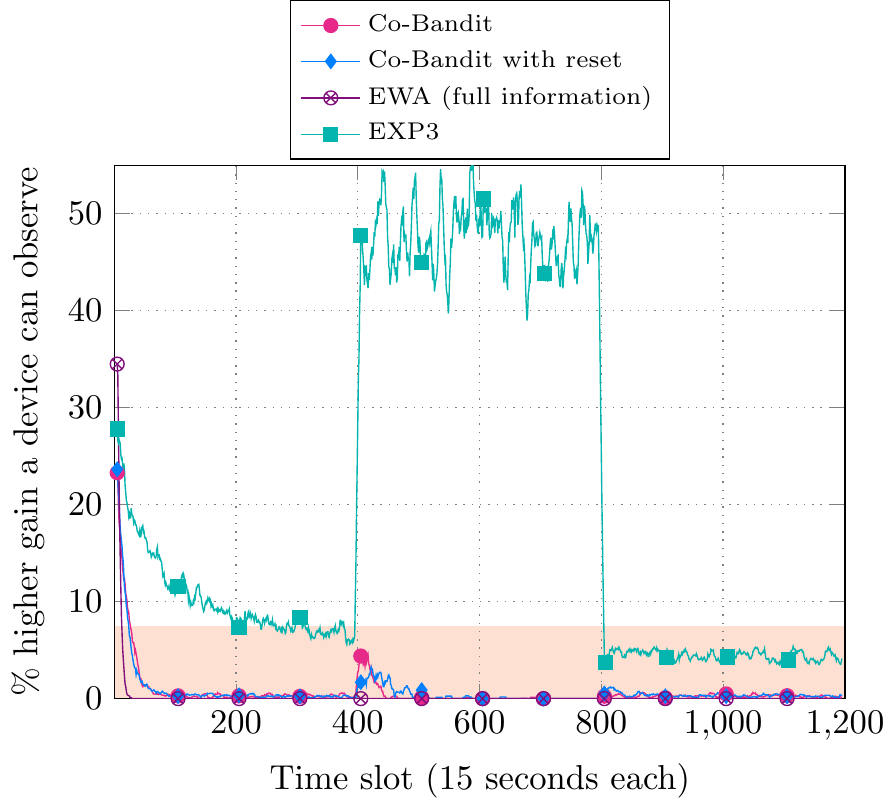}
\end{center}
\caption{Average distance to Nash equilibrium (\% higher gain any device would have observed, compared to its current gain, if the algorithm was at Nash equilibrium) when 10 devices join the service are at t = 401 and leave at the end of t = 800 --- shaded region represents $\epsilon$-equilibrium, where $\epsilon = 7.5$.}
\label{figure:dynamic_setting_2_distanceToNashEquilibrium}
\end{figure}

\noindent\textbf{Devices joining and leaving the service area.}
    We consider the same set of networks as in the previous dynamic setting. All devices have access to the same networks and can hear each other. But, 10 of them join the service area at the beginning of t = 401 and leave at the end of t = 800. Figure \ref{figure:dynamic_setting_2_distanceToNashEquilibrium} shows that Co-Bandit dynamically adapts to the changes, and far outperforms EXP3. Both Co-Bandit and EWA perform better in this setting compared to the previous dynamic setting. They are already stable at the optimal state when the devices join at t = 401. These devices fit into the setting, without causing much disruption. When they leave at t = 800, the algorithms quickly revert back to the stable state they were at prior to the 10 devices joining.

\noindent\textbf{Mobile users moving around.}
    We now consider the service areas in Figure \ref{figure:heterogeneousNetworks}, where networks 1, 2, 3, 4 and 5 have data rates 16, 14, 22, 7 and 4 Mbps, respectively. Initially there are 10 devices (1 - 10) at the food court, 5 devices (11 - 15) at the study area and 5 devices (16 - 20) at the bus stop. 8 devices (1 - 8) from the food court move to the study area at the beginning of $t = 401$ and eventually reach the bus stop at the start of $t = 801$. We assume that all devices in a service area can hear each other, e.g., all devices at the food court can hear each other but cannot hear devices at the study area.

\begin{figure}[!htb]
\begin{center}
\includegraphics %[scale=0.78]
{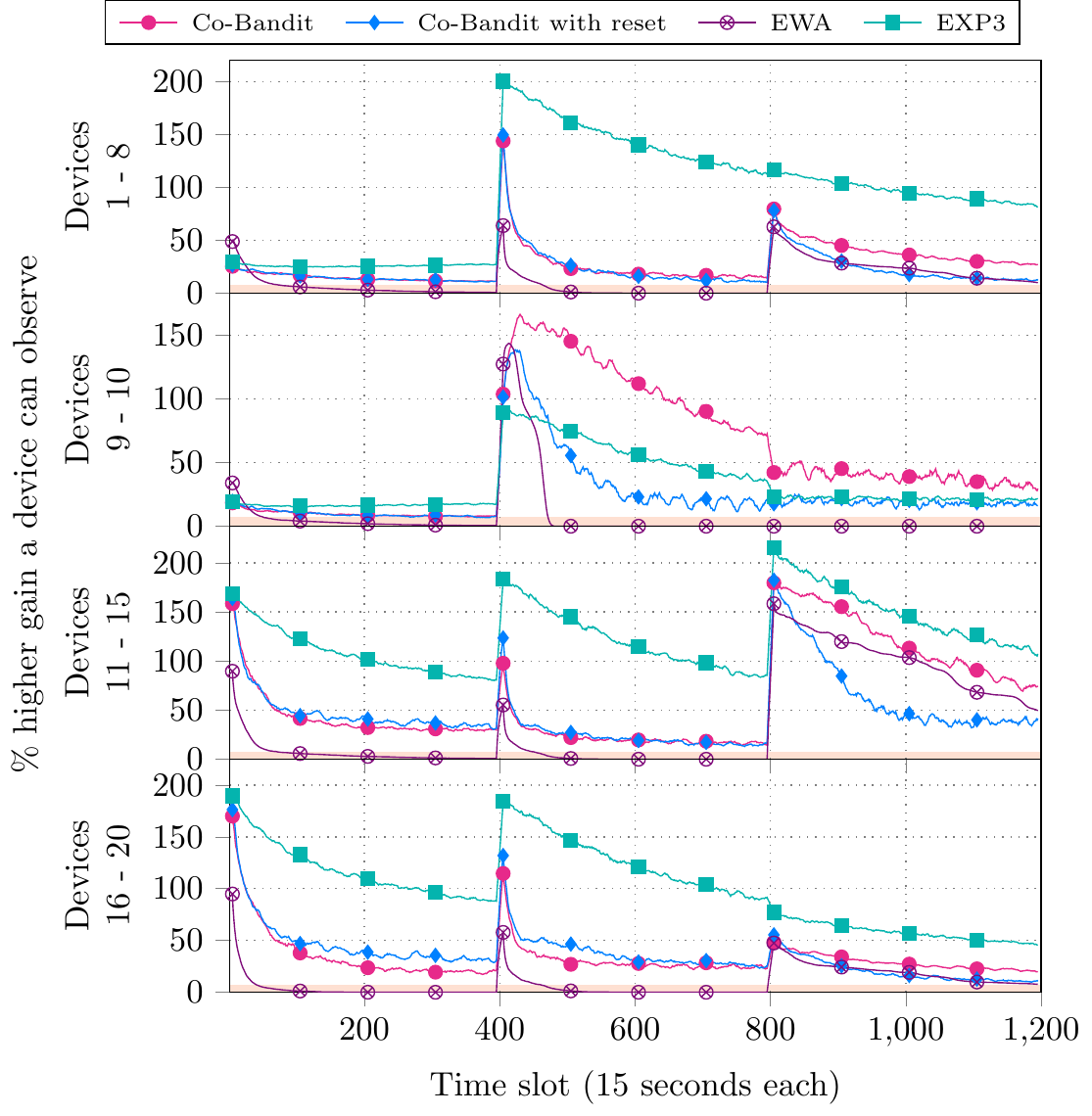}
\end{center}
\caption{Average distance to Nash equilibrium (\% higher gain any device would have observed, compared to its current gain, if the algorithm was at Nash equilibrium) in the setting shown in Figure \ref{figure:heterogeneousNetworks} with 8 users moving from area A to C through B --- shaded region represents $\epsilon$-equilibrium, where $\epsilon = 7.5$.}
\label{figure:mobility_setting_distanceToNashEquilibrium}
\end{figure}

Figure \ref{figure:mobility_setting_distanceToNashEquilibrium} shows the performance of each algorithm for devices in each area and those moving across areas, separately.
% 
% A - 1, 2, 3
% B - 1, 3, 4, 5
% C - 1, 4, 5
% NE:-
% phase 1: 5,5,7,2,1
% phase 2: 6,2,9,2,1
% phase 3: 8,2,5,3,2
% 
We observe similar behaviour as in static settings during the first 400 time slots for all algorithms. However, the distance for Co-Bandit is higher for devices at the study area and bus stop. This is due to the cost of exploring networks unheard of (in particular network 5, common to both areas, to which a single device is associated at optimal state --- thus, the probability of hearing about it is low). 
As devices move across service areas, they disrupt the setting, but eventually Co-Bandit adapts accordingly and outperforms EXP3. Yet, we observe more exploration in this setting because of our assumption that devices in different areas do not hear from each other. 
For example, devices 9 and 10 have to associate with network 2 as from $t = 401$; they will only be able to hear from each other at that time and will keep exploring networks 1 and 2 with uniform probability every 32 time slots. The minimal reset improves the rate of adaptation to changes.
% Some times Smart EXP3 performs slightly better. 
% Devices using EWA at the study area seems to adapt, but extremely slowly, when devices leave their area at the end of $t = 800$.

% \color{red} must re-run simulation without reset \color{black}

\section{Other related work} \label{section:relatedWorks}
    In this section, we discuss state-of-art wireless network selection approaches, and relevant work done on cooperative bandit, the use of graph to model communication network and delayed feedback.

% cooperative network selection
A significant amount of work consider the use of multiple wireless networks, such as Multinet \cite{chandra2004multinet}, and MPTCP \cite{ford2013tcp}. Yet, 
% and Coolspot \cite{pering2006coolspots}. However, Coolspot focuses on saving energy by switching between WiFi and Bluetooth, and 
identifying the optimal network is crucial for good performance 
% even in Multinet and MPTCP 
\cite{deng2014wifi}. A number of centralized wireless network selection approaches \cite{aryafar2017max, bejerano2004fairness, mishra2006client, sui2016characterizing} have been proposed. But, they are not scalable and are limited to managed networks. Several distributed solutions have been presented. Some require coordination from APs \cite{kauffmann2007measurement}. Others require global knowledge \cite{niyato2009dynamics,aryafar2013rat, monsef2015convergence}, or availability of some information \cite{zhu2010network,cheung2017congestion}, or assume a stochastic bandit setting \cite{wu2016traffic}. They all have some limitations. In our prior work, we proposed Smart EXP3 \cite{appavoo2018shrewd}, a bandit algorithm with good theoretical and practical performance (far better than EXP3). Here, we consider cooperation for an improved rate of stabilization.
A cooperative approach to network selection was considered in \cite{deng2014all}, where devices estimate and share properties of their networks to other devices associated to the same network. We consider sharing across networks, although within a range.

The closest to our work is \cite{cesa2016delay}, in which the authors proposed an algorithm that allows agents to share feedback with their neighbors relying on a communication network modeled as a graph. Feedback is used as soon as it is received and those older than some threshold are dropped. However, they consider an abstract problem and an abstract graph, and give an average welfare regret bound that relies on combinatorial graph properties. We solve the wireless network selection problem and consider a random graph based on communication pattern in a dynamic wireless network environment. As such, we provide a stronger per-device regret bound, which is better suited to the wireless network setting and highlights the impact of varying the amount of cooperation and delay. In addition, we consider stabilization of our algorithm in a multi-agent setting.

Graph structured feedback has been studied in numerous other work.
In \cite{alon2017nonstochastic}, relationship between actions is modelled using a time-changing directed graph, and the agent gets instant feedback about its chosen action and related actions. A considerable amount of work has considered networks of cooperative stochastic bandits using dynamic peer-to-peer random network \cite{szorenyi2013gossip}, fixed communication graph \cite{landgren2016distributed}, and social network \cite{kolla2018collaborative}. Cooperative contextual bandit is studied in \cite{tekin2015distributed} where each agent can select an action or request another agent to select it (with a cost). All the work focus on minimizing (in most cases, average) regret. A significant amount of work has considered delayed feedback, focusing on its impact on regret. Several work assumed full information \cite{weinberger2002delayed, mesterharm2005line, mesterharm2007improving, joulani2016delay, quanrud2015online} or stochastic settings \cite{dudik2011efficient}, considering both fixed and variable delay. The traditional sub-optimal approaches to deal with delayed feedback are to (a) use multiple instances of a non-delayed algorithm \cite{weinberger2002delayed, cesa2016delay, joulani2013online}, or (b) wait for all feedback to be received before taking the next decision.

\section{Conclusion} \label{section:conclusion}
    EXP3, a leading multi-armed bandit algorithm, has excellent theoretical properties but takes an unacceptable amount of time to stabilize in practice. 
Full information setting requires support from network service providers, which may be infeasible. Hence, we consider a spectrum of settings between full information and bandit settings with cooperation among devices.
% already have many cooperative applications
We have proposed Co-Bandit, a novel cooperative bandit algorithm and evaluated its performance in dynamic wireless network settings, where a mobile device has to select the optimal wireless network for good performance. Empirical results show that it far outperforms EXP3. A little cooperation among devices, even when feedback is received with a delay, can significantly enhance performance and the rate of learning.
% While the amount of data shared occasionally may be a concern, data may be compressed before sharing. 

As future work, we intend to (a) evaluate Co-Bandit in real-world settings, (b) further enhance its performance through features of Smart EXP3, e.g., cater for stability when more than one network offers the same bit rate, (c) consider settings with dishonest devices, i.e., settings where devices lie about their observations, and (d) apply it to other resource selection problems requiring fast stabilization.
% 
% future works: real-world evaluation
% need to propagate the probabilities????? any way to improve this?
% amount of communication may be a concern, but we can have compression techniques to combine data into more compact format; can we use network coding?
% how to estimate number of devices associated to cellular network? or how to estimate what I can get from the network...
% sharing using e.g. Bluetooth
    
\bibliographystyle{plain}
\bibliography{references}

\begin{thebibliography}{10}

\bibitem{alon2017nonstochastic}
Noga Alon, Nicolo Cesa-Bianchi, Claudio Gentile, Shie Mannor, Yishay Mansour,
  and Ohad Shamir.
\newblock Nonstochastic multi-armed bandits with graph-structured feedback.
\newblock {\em SIAM Journal on Computing}, 46(6):1785--1826, 2017.

\bibitem{appavoo2018shrewd}
Anuja~Meetoo Appavoo, Seth Gilbert, and Kian-Lee Tan.
\newblock Shrewd selection speeds surfing: Use smart exp3!
\newblock In {\em 2018 IEEE 38th International Conference on Distributed
  Computing Systems (ICDCS)}, pages 188--199. IEEE, 2018.

\bibitem{arp_scan}
arp scan.
\newblock The arp scanner - linux man page.
\newblock \url{https://linux.die.net/man/1/arp-scan}, 2018.

\bibitem{aryafar2017max}
E.~Aryafar, A.~Keshavarz-Haddad, C.~Joe-Wong, and M.~Chiang.
\newblock Max-min fair resource allocation in hetnets: Distributed algorithms
  and hybrid architecture.
\newblock In {\em ICDCS, 2017}, pages 857--869. IEEE, 2017.

\bibitem{aryafar2013rat}
E.~Aryafar, A.~Keshavarz-Haddad, M.l Wang, and M.~Chiang.
\newblock Rat selection games in hetnets.
\newblock In {\em INFOCOM}, pages 998--1006. IEEE, 2013.

\bibitem{auer2002nonstochastic}
P.~Auer, N.~Cesa-Bianchi, Y.~Freund, and R.~E. Schapire.
\newblock The nonstochastic multiarmed bandit problem.
\newblock {\em SIAM Journal on Computing}, 32(1):48--77, 2002.

\bibitem{bejerano2004fairness}
Y.~Bejerano, S-J. Han, and L.~E. Li.
\newblock Fairness and load balancing in wireless lans using association
  control.
\newblock In {\em MobiCom}, pages 315--329. ACM, 2004.

\bibitem{cesa2016delay}
Nicolo Cesa-Bianchi, Claudio Gentile, Yishay Mansour, and Alberto Minora.
\newblock Delay and cooperation in nonstochastic bandits.
\newblock {\em Journal of Machine Learning Research}, 49:605--622, 2016.

\bibitem{cesa2006prediction}
Nicolo Cesa-Bianchi and G{\'a}bor Lugosi.
\newblock {\em Prediction, learning, and games}.
\newblock Cambridge university press, 2006.

\bibitem{chandra2004multinet}
R.~Chandra and P.~Bahl.
\newblock Multinet: Connecting to multiple ieee 802.11 networks using a single
  wireless card.
\newblock In {\em INFOCOM}, volume~2, pages 882--893. IEEE, 2004.

\bibitem{cheung2017congestion}
M.~H. Cheung, F.~Hou, J.~Huang, and R.~Southwell.
\newblock Congestion-aware distributed network selection for integrated
  cellular and wi-fi networks.
\newblock {\em arXiv preprint arXiv:1703.00216}, 2017.

\bibitem{deng2014wifi}
S.~Deng, R.~Netravali, A.~Sivaraman, and H.~Balakrishnan.
\newblock Wifi, lte, or both?: Measuring multi-homed wireless internet
  performance.
\newblock In {\em IMC}, pages 181--194. ACM, 2014.

\bibitem{deng2014all}
S.~Deng, A.~Sivaraman, and H.~Balakrishnan.
\newblock All your network are belong to us: A transport framework for mobile
  network selection.
\newblock In {\em HotMobile}. ACM, 2014.

\bibitem{dudik2011efficient}
Miroslav Dudik, Daniel Hsu, Satyen Kale, Nikos Karampatziakis, John Langford,
  Lev Reyzin, and Tong Zhang.
\newblock Efficient optimal learning for contextual bandits.
\newblock {\em arXiv preprint arXiv:1106.2369}, 2011.

\bibitem{fitter}
Fitter.
\newblock A tool to fit data to many distributions and best one(s) for python.
\newblock \url{https://pypi.python.org/pypi/fitter}, 2016.

\bibitem{ieee80211}
IEEE~Standard for Information~technology (2009).
\newblock Local and metropolitan area networks– specific requirements–part
  11.
\newblock Technical report, IEEE, 2009.

\bibitem{ford2013tcp}
A.~Ford, C.~Raiciu, M.~Handley, and O.~Bonaventure.
\newblock Tcp extensions for multipath operation with multiple addresses.
\newblock Technical report, Ford, 2013.

\bibitem{gessner2012umts}
C.~Gessner, A.~Roessier, and M.~Kottkamp.
\newblock Umts long term evolution (lte)--technology introduction application
  note., 2012.

\bibitem{googlemap}
Business Insider.
\newblock {Here's how Google Maps knows when there is traffic.}, 2017.
\newblock \url{https://tinyurl.com/ycymy4wp}, accessed 2019-01-07.

\bibitem{joulani2013online}
Pooria Joulani, Andras Gyorgy, and Csaba Szepesv{\'a}ri.
\newblock Online learning under delayed feedback.
\newblock In {\em International Conference on Machine Learning}, pages
  1453--1461, 2013.

\bibitem{joulani2016delay}
Pooria Joulani, Andr{\'a}s Gy{\"o}rgy, and Csaba Szepesv{\'a}ri.
\newblock Delay-tolerant online convex optimization: Unified analysis and
  adaptive-gradient algorithms.
\newblock In {\em AAAI}, volume~16, pages 1744--1750, 2016.

\bibitem{kauffmann2007measurement}
B.~Kauffmann, F.~Baccelli, A.~Chaintreau, V.~Mhatre, K.~Papagiannaki, and
  C.~Diot.
\newblock Measurement-based self organization of interfering 802.11 wireless
  access networks.
\newblock In {\em INFOCOM 2007}, pages 1451--1459. IEEE, 2007.

\bibitem{kleinberg2009multiplicative}
R.~Kleinberg, G.~Piliouras, and E.~Tardos.
\newblock Multiplicative updates outperform generic no-regret learning in
  congestion games.
\newblock In {\em ACM STOC}, pages 533--542. ACM, 2009.

\bibitem{kolla2018collaborative}
Ravi~Kumar Kolla, Krishna Jagannathan, and Aditya Gopalan.
\newblock Collaborative learning of stochastic bandits over a social network.
\newblock {\em IEEE/ACM Transactions on Networking (TON)}, 26(4):1782--1795,
  2018.

\bibitem{landgren2016distributed}
Peter Landgren, Vaibhav Srivastava, and Naomi~Ehrich Leonard.
\newblock On distributed cooperative decision-making in multiarmed bandits.
\newblock In {\em Control Conference (ECC), 2016 European}, pages 243--248.
  IEEE, 2016.

\bibitem{maghsudi2013relay}
S.~Maghsudi and S.~Stanczak.
\newblock Relay selection with no side information: An adversarial bandit
  approach.
\newblock In {\em WCNC}, pages 715--720. IEEE, 2013.

\bibitem{mesterharm2005line}
Chris Mesterharm.
\newblock On-line learning with delayed label feedback.
\newblock In {\em International Conference on Algorithmic Learning Theory},
  pages 399--413. Springer, 2005.

\bibitem{mesterharm2007improving}
Chris Mesterharm.
\newblock {\em Improving on-line learning}.
\newblock PhD thesis, Rutgers University-Graduate School-New Brunswick, 2007.

\bibitem{mishra2006client}
A.~Mishra, V.~Brik, S.~Banerjee, A.~Srinivasan, and W.~A. Arbaugh.
\newblock A client-driven approach for channel management in wireless lans.
\newblock In {\em Infocom}, 2006.

\bibitem{monsef2015convergence}
E~Monsef, A.~Keshavarz-Haddad, E.~Aryafar, J.~Saniie, and M.~Chiang.
\newblock Convergence properties of general network selection games.
\newblock In {\em INFOCOM}, pages 1445--1453. IEEE, 2015.

\bibitem{nisan2007algorithmic}
Noam Nisan, Tim Roughgarden, Eva Tardos, and Vijay~V Vazirani.
\newblock {\em Algorithmic game theory}, volume~1.
\newblock Cambridge University Press Cambridge, 2007.

\bibitem{niyato2009dynamics}
D.~Niyato and E.~Hossain.
\newblock Dynamics of network selection in heterogeneous wireless networks: An
  evolutionary game approach.
\newblock {\em TVT}, 58(4):2008--2017, 2009.

\bibitem{quanrud2015online}
Kent Quanrud and Daniel Khashabi.
\newblock Online learning with adversarial delays.
\newblock In {\em Advances in Neural Information Processing Systems}, pages
  1270--1278, 2015.

\bibitem{rosenthal1973class}
R.~W. Rosenthal.
\newblock A class of games possessing pure-strategy nash equilibria.
\newblock {\em International Journal of Game Theory}, 2(1):65--67, 1973.

\bibitem{simpy}
SimPy.
\newblock {SimPy - Event discrete simulation for Python}, 2016.
\newblock \url{https://simpy.readthedocs.io/}, accessed 2018-19-12.

\bibitem{sui2016characterizing}
K.~Sui, M.~Zhou, D.~Liu, M.~Ma, D.~Pei, Y.~Zhao, Z.~Li, and T.~Moscibroda.
\newblock Characterizing and improving wifi latency in large-scale operational
  networks.
\newblock In {\em MobiSys}, pages 347--360. ACM, 2016.

\bibitem{szorenyi2013gossip}
Bal{\'a}zs Sz{\"o}r{\'e}nyi, R{\'o}bert Busa-Fekete, Istv{\'a}n Heged{\H{u}}s,
  R{\'o}bert Orm{\'a}ndi, M{\'a}rk Jelasity, and Bal{\'a}zs K{\'e}gl.
\newblock Gossip-based distributed stochastic bandit algorithms.
\newblock In {\em Journal of Machine Learning Research Workshop and Conference
  Proceedings}, volume~2, pages 1056--1064. International Machine Learning
  Societ, 2013.

\bibitem{tekin2011performance}
C.~Tekin and M.~Liu.
\newblock Performance and convergence of multi-user online learning.
\newblock In {\em GAMENETS}, pages 321--336. Springer, 2011.

\bibitem{tekin2015distributed}
Cem Tekin and Mihaela van~der Schaar.
\newblock Distributed online learning via cooperative contextual bandits.
\newblock {\em IEEE Transactions on Signal Processing}, 63(14):3700--3714,
  2015.

\bibitem{waze}
Waze.
\newblock {Get the best route, every day, with real–time help from other
  drivers.}, 2019.
\newblock \url{https://www.waze.com}, accessed 2019-01-07.

\bibitem{weinberger2002delayed}
Marcelo~J Weinberger and Erik Ordentlich.
\newblock On delayed prediction of individual sequences.
\newblock In {\em Information Theory, 2002. Proceedings. 2002 IEEE
  International Symposium on}, page 148. IEEE, 2002.

\bibitem{wu2016traffic}
Q.~Wu, Z.~Du, P.~Yang, Y.-D. Yao, and J.~Wang.
\newblock Traffic-aware online network selection in heterogeneous wireless
  networks.
\newblock {\em TVT}, 65(1):381--397, 2016.

\bibitem{zhu2010network}
K.~Zhu, D.~Niyato, and P.~Wang.
\newblock Network selection in heterogeneous wireless networks: Evolution with
  incomplete information.
\newblock In {\em WCNC}, pages 1--6. IEEE, 2010.

\end{thebibliography}

\appendices
\section{Proof of upper bound on weak regret} \label{appendix:regretBound}
    We present some facts derived from definitions, and Lemmas that are used in the proofs. Lemma \ref{lemma:sum_networks_p_q} upper bounds the expected sum, for all the networks $i \in \mathcal{K}$, of the ratio of any device $j$'s probability of choosing network $i$ at time $t - \tilde{t}$ to the probability that at least one of the devices it heard from (including itself) chooses network $i$ at time $t - \tilde{t}$.
Moreover, bounding the extent to which the probability distribution can drift in $d$ time slots (due to the delayed feedback), plays an important role in controlling regret. Lemmas \ref{lemma:upper_bound_probability_drift} and \ref{lemma:lower_bound_probability_drift} control its evolution, by lower bounding and upper bounding the drift, respectively.\\

\setcounter{equation}{0} % reset numbering from previous appendix

%%%%% fact 1: value of b_1, i.e. probability of having an edge between two vertices, given delayed feedback
% 
\noindent\textbf{Fact 1.}
\noindent Given that feedback received from neighbors are forwarded, as time elapses, the probability of receiving a particular observation increases (more devices have the information and are broadcasting it). Here, we compute $b_{\tilde{t}}$, the probability of learning a device's observation made at time $t - \tilde{t}$, for $\tilde{t} \ge 0$, i.e. within a delay of $\tilde{t}$ time slots. The length of the path between the two vertices (devices) involved can be up to $\tilde{t} + 1$.\\

\noindent Let Y be a discrete random variable that represents the length of a directed path between two vertices in the communication graph in a particular time slot, and $b_0$ be the probability of (directly) hearing from a device. \\

\noindent Then,
\begin{align}
&P(Y = 1) = b_0 \nonumber \\
&P(Y = 2) = \frac{(n - 2)!}{(n - 2 - 1)!}\ {b_0}^2 \text{ , considering all permutations of path without repetition of vertices} \nonumber \\
&P(Y = 3) = \frac{(n - 2)!}{(n - 2 - 2)!}\ {b_0}^3 \nonumber \\
&... \nonumber \\
&P(Y = \tilde{t} + 1) = \frac{(n - 2)!}{(n - 2 - \tilde{t})!}\ {b_0}^{\tilde{t}+1} \text{, when } n \ge \tilde{t} + 2 \nonumber
\end{align}

\noindent When the delay $\tilde{t}$ increases beyond $n - 2$, the probability remains the same since all paths the message can take have already been considered. \\

\noindent Therefore,
\begin{align}
b_{\tilde{t}} 
&= 1 - \prod_{t' = 0}^{\tilde{t}} \left(1 - min\left\{\frac{(n-2)!}{max\{(n-2-t'), 0\}!}\ {b_0}^{t'+1}, 1\right\}\right)
\label{fact:prob_hearing_with_delay}
\end{align}

%%%%% fact 2: expectation of I/q
% 
\noindent\textbf{Fact 2.}
\begin{align}
\mathbb{E}_{t-\tilde{t}}\left[\frac{I_{i,j}(t-\tilde{t}, t)}{q_{i,j}(t-\tilde{t}, t)}\right] &= \frac{1}{q_{i,j}(t-\tilde{t}, t)}\ \mathbb{E}_{t-\tilde{t}}\left[I_{i,j}(t-\tilde{t}, t)\right] \nonumber \\
&= \frac{1}{q_{i,j}(t-\tilde{t}, t)} \cdot q_{i,j}(t-\tilde{t}, t) \text{, by definition of $I_{i,j}(t-\tilde{t}, t)$} \nonumber \\
&= 1 
\label{fact:expectation_I_over_q}
% \label{proof:regret_fact_1}
\end{align}

%%%%% fact 3: expectation of loss estimate
% 
\noindent\textbf{Fact 3.}
\begin{align}
\mathbb{E}_{t-\tilde{t}}\left[\widehat{l_{i,j}}(t)\right] 
&= \mathbb{E}_{t-\tilde{t}}\left[\dfrac{1}{d' + 1}\ \sum\limits_{\tilde{t}=0}^{d'} \dfrac{l_{i,j}(t-\tilde{t}, t)}{q_{i,j}(t-\tilde{t},t)}\ I_{i,j}(t-\tilde{t},t)\right] \text{, from the loss estimate rule} \nonumber \\
&= \dfrac{1}{d' + 1}\ \sum\limits_{\tilde{t}=0}^{d'} l_{i,j}(t-\tilde{t}, t) \ \mathbb{E}_{t-\tilde{t}}\left[\frac{I_{i,j}(t-\tilde{t},t)}{q_{i,j}(t-\tilde{t},t)}\right] \nonumber \\
&= \dfrac{1}{d'+1}\ \sum\limits_{\tilde{t}=0}^{d'} l_{i,j}(t-\tilde{t}, t) \text{, using Fact \ref{fact:expectation_I_over_q}} \nonumber \\
&= l_{i,j}(t) 
\label{fact:expectation_loss_estimate}
\end{align}
\\
\begin{lemma}
Given $n$ active mobile devices, and
$k$ wireless networks in the service area,
for any device $j \in \mathcal{N}$, $b_0 \in[0,1 - e^{-1}]$, $x = d$ at time $t$, and any time $t \ge 1$, we can say that
\begin{align}
\mathbb{E}\left[\sum_{i \in \mathcal{K}} \frac{p_{i,j}(t-\tilde{t})}{q_{i,j}(t-\tilde{t}, t)}\right]
\le \frac{1}{\max\{\frac{1}{k}, b_0\}} \nonumber
\end{align}
\label{lemma:sum_networks_p_q}
\end{lemma}
% lemma 1
\begin{IEEEproof}[\textbf{Proof}]
\setcounter{equation}{0} % reset numbering from previous appendix
$q_{i,j}(t-\tilde{t}, t)$ is the probability that the quality of network $i$ at time $t - \tilde{t}$ is known to device $j$ by now (by the current time $t$), whether by exploring it at time $t - \tilde{t}$ or by hearing about it from neighbor(s) over the past $\tilde{t}$ time slots.

\noindent By definition,
\begin{align}
q_{i,j}(t-\tilde{t}, t) 
&= 1 - \prod_{j' \in H_j(t-\tilde{t},t)} (1 - p_{i,j'}(t-\tilde{t})) \nonumber \\
&= 1 - \left(1 - p_{i,j}(t-\tilde{t})\right) \underbrace{\prod_{j' \in H_j(t-\tilde{t},t) - \{j\}} \left(1 - p_{i,j'}(t-\tilde{t})\right)}_\text{\parbox{4cm}{\centering probability no device shared\\[-4pt] details about network $i$}} \nonumber
\end{align}

\noindent We now compute the probability that some device has shared details about network $i$. We assume that $t \ge d + 1$ and $x = d$. We consider two cases, namely (a) at least one device $j'$ explored the network as it was unheard of for $x$ time slots, and (b) some device $j'$ was associated to it and shared its observation; this observation may have been propagated by other devices if $\tilde{t} > 0$.\\

\noindent In the first case, a device $j'$ selects network $i$ with probability $\frac{1}{n}$ and shares its observation with probability $1$. In this case, the probability that some device explores and shares about network $i$ is given as
$$1 - \left(1 - \frac{1}{n}\right)^{n-1} \approx 1 - e^{-1}$$

\noindent In the second case, some device $j'$ has selected it and device $j$ will hear from $j'$ about $i$ with probability $b_{\tilde{t}}$.\\

\noindent Hence, the probability that some device has shared details about network $i$ 
$\ge \min\{1 - e^{-1}, b_{\tilde{t}}\}$, where $b_{\tilde{t}} \ge b_0$.

\noindent Thus, we can say that
\begin{align}
q_{i,j}(t-\tilde{t}, t) 
&\ge 1 - \left(1 - p_{i,j}(t-\tilde{t})\right) \left( 1 - \min\{1 - e^{-1}, b_{0}\}\right) \nonumber \\
&\ge 1 - \left(1 - p_{i,j}(t-\tilde{t})\right) \left( 1 - b_{0}\right) \text{, as we assume that } b_0 \le 1 - e^{-1}\nonumber
\end{align}

\noindent Then,
\begin{align}
\mathbb{E}\left[\sum_{i \in \mathcal{K}} \frac{p_{i,j}(t - \tilde{t})}{q_{i,j}(t - \tilde{t},t)}\right]
&\le \mathbb{E}\left[\sum_{i \in \mathcal{K}} \frac{p_{i,j}(t - \tilde{t})}{1 - \left(1 - p_{i,j}(t-\tilde{t})\right) \left( 1 - b_{0}\right)}\right] \nonumber \\
&\le \sum_{i \in \mathcal{K}} \frac{\frac{1}{k}}{1 - \left(1 - \frac{1}{k}\right) \left( 1 - b_{0}\right)} \nonumber \\
&\le \frac{1}{1 - \left(1 - \frac{1}{k}\right) \left( 1 - b_{0}\right)} \nonumber
\end{align}

\noindent We can say that 
$ \dfrac{1}{1 - \left(1 - \frac{1}{k}\right) \left( 1 - b_{0}\right)} 
\le \dfrac{1}{1 - \left(1 - \frac{1}{k}\right)}
\le \dfrac{1}{\frac{1}{k}}
$

\noindent We can also say that
$\dfrac{1}{1 - \left(1 - \frac{1}{k}\right) \left( 1 - b_{0}\right)} 
\le \dfrac{1}{1 - \left(1 - b_0\right)} 
\le \dfrac{1}{\frac{1}{b_0}} $

\noindent As such, $\dfrac{1}{1 - \left(1 - \frac{1}{k}\right) \left( 1 - b_{0}\right)} 
\le \dfrac{1}{\max\{\frac{1}{k}, b_0\}}$

\noindent Therefore, 
\begin{align}
\mathbb{E}\left[\sum_{i \in \mathcal{K}} \frac{p_{i,j}(t - \tilde{t})}{q_{i,j}(t - \tilde{t},t)}\right]
&\le \frac{1}{\max\{\frac{1}{k}, b_0\}} \nonumber
\end{align}
\noindent which concludes the proof.
\end{IEEEproof}
\BlankLine
\BlankLine

\begin{lemma} 
For any device $j \in \mathcal{N}$, each network $i \in \mathcal{K}$, any $t \ge  1 $ when $d \in \mathbb{Z}_{\ge 0}$, and $0 < \eta \le \dfrac{1}{ke(d+1)}$, 
\begin{align}
p_{i,j}(t+1) \le \left(1 + \frac{1}{d}\right) p_{i,j}(t) \nonumber
\end{align}
\label{lemma:upper_bound_probability_drift}
\end{lemma}
\begin{IEEEproof}[\textbf{Proof}]
We follow the proofs of Lemmas 1, 2 and 19 in \cite{cesa2016delay}. 
\setcounter{equation}{0} % reset numbering from previous appendix

\begin{align}
p_{i,j}(t+1) - p_{i,j}(t) 
&= p_{i,j}(t+1) - \frac{w_{i,j}(t)}{\sum\limits_{m=1}^{k}w_{m,j}(t)} \text{, using the probability update rule}
\label{proof:lemma3_1}
\end{align}

\noindent From the weight update rule, we have
\begin{align}
w_{i,j}(t + 1) &= \dfrac{w_{i,j}(t)\ exp(-\eta\ \widehat{l_{i,j}}(t))}{\underset{m \in \mathcal{K}}{\max}\{w_{m,j}(t)\ exp(-\eta\ \widehat{l_{m,j}}(t))\}} \nonumber
\end{align}

\begin{align}
w_{i,j}(t) &= \dfrac{w_{i,j}(t + 1)\ \underset{m \in \mathcal{K}}{\max}\{w_{m,j}(t)\ exp(-\eta\ \widehat{l_{m,j}}(t))\}}{exp(-\eta\ \widehat{l_{i,j}}(t))} \nonumber \\
&\ge w_{i,j}(t + 1)\ \underset{m \in \mathcal{K}}{\max}\{w_{m,j}(t)\ exp(-\eta\ \widehat{l_{m,j}}(t))\} \text{, as } exp(-\eta\ \widehat{l_{i,j}}(t)) \le 1 \nonumber
\end{align}

\noindent Combining this with (\ref{proof:lemma3_1}), we get
\begin{align}
p_{i,j}(t+1) - p_{i,j}(t) 
&\le p_{i,j}(t+1) - \frac{w_{i,j}(t + 1)\ \underset{m \in \mathcal{K}}{\max}\{w_{m,j}(t)\ exp(-\eta\ \widehat{l_{m,j}}(t))\}}{\sum\limits_{m=1}^{k}w_{m,j}(t)}
\label{proof:lemma3_2}
\end{align}

\noindent From the probability update rule, we have
\begin{align}
p_{i,j}(t+1) &= \frac{w_{i,j}(t+1)}{\sum\limits_{m=1}^{k}w_{m,j}(t+1)} \nonumber \\
w_{i,j}(t+1) &= \sum\limits_{m=1}^{k}w_{m,j}(t+1)\ p_{i,j}(t+1) 
% \label{proof:lemma3_3}
\end{align}

\noindent Combining this with (\ref{proof:lemma3_2}), we have
\begin{align}
p_{i,j}(t+1) - p_{i,j}(t) 
&\le p_{i,j}(t+1) - \frac{\sum\limits_{m=1}^{k}w_{m,j}(t+1) \ p_{i,j}(t+1) \ \underset{y \in \mathcal{K}}{\max}\{w_{y,j}(t)\ exp(-\eta\ \widehat{l_{y,j}}(t))\}}{\sum\limits_{m=1}^{k}w_{m,j}(t)}  \nonumber \\
&\le p_{i,j}(t+1) - \frac{\sum\limits_{m=1}^{k} \frac{w_{m,j}(t)\ exp(-\eta\ \widehat{l_{m,j}}(t))}{\underset{y \in \mathcal{K}}{\max}\{w_{y,j}(t)\ exp(-\eta\ \widehat{l_{y,j}}(t))\}} \ p_{i,j}(t+1) \ \underset{y \in \mathcal{K}}{\max}\{w_{y,j}(t)\ exp(-\eta\ \widehat{l_{y,j}}(t))\}}{\sum\limits_{m=1}^{k}w_{m,j}(t)} \text{ , } \nonumber \\
&\hspace{5mm} \text{using the weight update rule} \nonumber \\ 
&\le p_{i,j}(t+1) - \frac{\sum\limits_{m=1}^{k} w_{m,j}(t)\ exp(-\eta\ \widehat{l_{m,j}}(t))\ p_{i,j}(t+1) }{\sum\limits_{m=1}^{k}w_{m,j}(t)} \nonumber \\
&\le p_{i,j}(t+1)\left(1 - \frac{\sum\limits_{m=1}^{k} w_{m,j}(t)\ exp(-\eta\ \widehat{l_{m,j}}(t))}{\sum\limits_{m=1}^{k}w_{m,j}(t)} \right) 
\label{proof:lemma3_4}
\end{align}

\noindent From the probability update rule, we have 
\begin{align}
p_{i,j}(t) &= \frac{w_{i,j}(t)}{\sum\limits_{y=1}^{k} w_{y,j}(t)} \nonumber \\
w_{i,j}(t) &= \sum\limits_{y=1}^{k} w_{y,j}(t)\ p_{i,j}(t) \nonumber
% w_{m,j}(t) 
% &= \dfrac{\sum\limits_{y=1}^{k}w_{y,j}(t) \left(p_{m,j}(t) - \frac{\gamma}{k}\right)}{1-\gamma} \nonumber \\
% &\le \sum\limits_{y=1}^{k}w_{y,j}(t) \left(p_{m,j}(t) - \frac{\gamma}{k}\right) \nonumber
\end{align}

\noindent Combining this with (\ref{proof:lemma3_4}), we get
\begin{align}
p_{i,j}(t+1) - p_{i,j}(t) 
&\le p_{i,j}(t+1)\left(1 - \frac{\sum\limits_{m=1}^{k} \sum\limits_{y=1}^{k} w_{y,j}(t)\ p_{m,j}(t)\ exp(-\eta\ \widehat{l_{m,j}}(t))}{\sum\limits_{m=1}^{k}w_{m,j}(t)} \right) \nonumber \\
&\le p_{i,j}(t+1)\left(1 - \sum\limits_{m=1}^{k}  p_{m,j}(t)\ exp(-\eta\ \widehat{l_{m,j}}(t)) \right) \nonumber \\
&\le p_{i,j}(t+1)\left(\sum\limits_{m=1}^{k}  p_{m,j}(t) - \sum\limits_{m=1}^{k}  p_{m,j}(t)\ exp(-\eta\ \widehat{l_{m,j}}(t)) \right) \nonumber \\
&\le p_{i,j}(t+1) \sum\limits_{m=1}^{k} p_{m,j}(t) \left(1 - exp(-\eta\ \widehat{l_{m,j}}(t)) \right) \nonumber \\
&\le p_{i,j}(t+1) \sum\limits_{m=1}^{k} p_{m,j}(t) \left(\eta \ \widehat{l_{m,j}}(t)\right) \text{, as } 1 - e^{-x} \le x\nonumber \\
&\le \eta \ p_{i,j}(t+1) \sum\limits_{m=1}^{k} p_{m,j}(t) \ \widehat{l_{m,j}}(t)
\label{proof:lemma3_5}
\end{align}

\noindent We now upper bound $\sum_{m=1}^{k} p_{m,j}(t)\ \widehat{l_{m,j}}(t) $ by following an inductive argument similar to the ones in the proofs of Lemmas 2 and 19 in \cite{cesa2016delay}. For simplicity, we assume that $t \ge d + 1$.
\begin{align}
\sum_{m=1}^{k} p_{m,j}(t)\ \widehat{l_{m,j}}(t) 
&= \sum_{m=1}^{k} p_{m,j}(t) \cdot \dfrac{1}{d+1} \sum\limits_{\tilde{t}=0}^{d} \frac{l_{m,j}(t-\tilde{t}, t)}{q_{m,j}(t-\tilde{t},t)}\ I_{m,j}(t-\tilde{t},t) \text{, using the loss estimate rule} \nonumber \\
& \le \dfrac{1}{d+1} \sum_{m=1}^{k} p_{m,j}(t)\ \sum\limits_{\tilde{t}=0}^{d} \frac{1}{q_{m,j}(t-\tilde{t},t)} \text{, since } l_{m,j}(t-\tilde{t},t)\ I_{m,j}(t-\tilde{t},t) \le 1\nonumber \\
& \le \dfrac{1}{d+1} \sum_{m=1}^{k}\ \sum\limits_{\tilde{t}=0}^{d} \left(1 + \dfrac{1}{d}\right)^{\tilde{t}}\ \frac{p_{m,j}(t-\tilde{t})}{q_{m,j}(t-\tilde{t},t)} \text{, by inductive hypothesis} \nonumber \\
& \le \sum_{m=1}^{k}\ \sum\limits_{\tilde{t}=0}^{d} \dfrac{1}{d+1} \left(1 + \dfrac{1}{d}\right)^{\tilde{t}} \text{, as } q_{m,j}(t-\tilde{t},t) \ge p_{m,j}(t-\tilde{t}) \nonumber \\
& \le \sum_{m=1}^{k}\ e \text{, since } e \text{ approximates } \left(1 + \dfrac{1}{d}\right)^{\tilde{t}} \text{ and } \sum\limits_{\tilde{t}=0}^{d} \dfrac{1}{d+1} = 1 \nonumber \\
&\le ke \nonumber
\end{align}

\noindent Combining this with (\ref{proof:lemma3_5}), when $d \in \mathbb{Z}_{\ge 0}$, and $\eta \le \dfrac{1}{ke(d+1)}$, we get
\begin{align}
p_{i,j}(t+1) - p_{i,j}(t) 
&\le \eta k e \ p_{i,j}(t+1) \nonumber \\
&\le \frac{1}{ke(d + 1)} \cdot  k e \ p_{i,j}(t+1) \nonumber \\
&\le \frac{1}{d + 1}\ p_{i,j}(t+1) \nonumber
\end{align}

\begin{align}
p_{i,j}(t+1) \left(1 - \dfrac{1}{d+1}\right) \le p_{i,j}(t) \nonumber 
\end{align}
\begin{align}
p_{i,j}(t+1) \left(\dfrac{d}{d+1}\right) \le p_{i,j}(t) \nonumber
\end{align}
\begin{align}
p_{i,j}(t+1) \le \left(1 + \dfrac{1}{d}\right) p_{i,j}(t) \nonumber
\end{align}

\noindent which concludes the proof.

\end{IEEEproof}

\begin{lemma} 
For any device $j \in \mathcal{N}$, each network $i \in \mathcal{K}$, and any $t \ge d + 1$  when $d \in \mathbb{Z}_{\ge 0}$, $\eta > 0$, we have
\begin{align}
p_{i,j}(t+1)-p_{i,j}(t) 
&\ge - \frac{e \eta}{d} \sum\limits_{\tilde{t}=0}^{d} \frac{p_{i,j}(t-\tilde{t})\ I_{i,j}(t-\tilde{t},t)}{q_{i,j}(t-\tilde{t},t)}\nonumber 
\end{align}
\label{lemma:lower_bound_probability_drift}
\end{lemma}
\begin{IEEEproof}[\textbf{Proof}]
We follow the proofs of Lemmas 1, 2 and 20 in \cite{cesa2016delay}. 
\setcounter{equation}{0} % reset numbering from previous appendix

\noindent From the probability update rule, we have
\begin{align}
p_{i,j}(t+1) &= \frac{w_{i,j}(t+1)}{\sum\limits_{m=1}^{k}w_{m,j}(t+1)} \nonumber
\end{align}

\noindent Therefore, we can say that
\begin{align}
p_{i,j}(t+1)-p_{i,j}(t) 
&= \dfrac{w_{i,j}(t+1)}{\sum\limits_{m=1}^{k}w_{m,j}(t+1)}  - p_{i,j}(t) \nonumber \\
&= \dfrac{\dfrac{w_{i,j}(t)\ exp(-\eta\ \widehat{l_{i,j}}(t))}{\underset{y \in \mathcal{K}}{\max}\{w_{y,j}(t)\ exp(-\eta\ \widehat{l_{y,j}}(t))\}}}{\sum\limits_{m=1}^{k}\ \dfrac{w_{m,j}(t)\ exp(-\eta\ \widehat{l_{m,j}}(t))}{\underset{y \in \mathcal{K}}{\max}\{w_{y,j}(t)\ exp(-\eta\ \widehat{l_{y,j}}(t))\}}} - p_{i,j}(t) \text{, using the weight update rule} \nonumber \\
&= \dfrac{w_{i,j}(t)\ exp(-\eta\ \widehat{l_{i,j}}(t))}{\sum\limits_{m=1}^{k}\ w_{m,j}(t)\ exp(-\eta\ \widehat{l_{m,j}}(t))} - p_{i,j}(t) 
\label{proof:lemma4_1}
\end{align}

\noindent From the probability update rule, we get
\begin{align}
p_{i,j}(t) &= \dfrac{w_{i,j}(t)}{\sum\limits_{y=1}^{k}w_{y,j}(t)}\nonumber \\
w_{i,j}(t) &= \sum\limits_{y=1}^{k}w_{y,j}(t)\ p_{i,j}(t)\nonumber
\end{align}

\noindent Combining this with (\ref{proof:lemma4_1}), we get
\begin{align}
p_{i,j}(t+1)-p_{i,j}(t) 
&= \dfrac{\sum\limits_{y=1}^{k}w_{y,j}(t)\ p_{i,j}(t)\ exp(-\eta\ \widehat{l_{i,j}}(t))}{\sum\limits_{m=1}^{k}\ \sum\limits_{y=1}^{k}w_{y,j}(t)\ p_{m,j}(t) \ exp(-\eta\ \widehat{l_{m,j}}(t))} - p_{i,j}(t) \nonumber \\
&= \dfrac{p_{i,j}(t) \ exp(-\eta\ \widehat{l_{i,j}}(t))}{\sum\limits_{m=1}^{k}\ p_{m,j}(t)\ exp(-\eta\ \widehat{l_{m,j}}(t))} - p_{i,j}(t) \nonumber \\
&\ge p_{i,j}(t) \ exp(-\eta\ \widehat{l_{i,j}}(t))  - p_{i,j}(t) \text{, as } exp(-\eta\ \widehat{l_{m,j}}(t)) \le 1 \text{, and } \sum\limits_{m=1}^{k}\ p_{m,j}(t) = 1 \nonumber \\
&\ge p_{i,j}(t) \left(exp(-\eta\ \widehat{l_{i,j}}(t)) - 1 \right) \nonumber \\
&\ge -\eta \ p_{i,j}(t)\ \widehat{l_{i,j}}(t) \text{, since } e^{-x} - 1 \ge -x 
\label{proof:lemma4_2}
\end{align}
\noindent We now upper bound $p_{i,j}(t)\ \widehat{l_{i,j}}(t)$, as in Lemma \ref{lemma:upper_bound_probability_drift}. For simplicity, we assume that $t \ge d + 1$.
\begin{align}
p_{i,j}(t)\ \widehat{l_{i,j}}(t) &= \frac{p_{i,j}(t)}{d+1} \sum\limits_{\tilde{t}=0}^{d} \frac{l_{i,j}(t-\tilde{t}, t)}{q_{i,j}(t-\tilde{t}, t)}\ I_{i,j}(t-\tilde{t}, t)  \text{, using the loss estimate rule} \nonumber \\
&\le \frac{1}{d+1} \sum\limits_{\tilde{t}=0}^{d} \frac{p_{i,j}(t)\ l_{i,j}(t-\tilde{t}, t)}{q_{i,j}(t-\tilde{t},t)}\ I_{i,j}(t-\tilde{t},t) \nonumber \\
&\le \frac{1}{d+1} \sum\limits_{\tilde{t}=0}^{d} \frac{p_{i,j}(t)\ }{q_{i,j}(t-\tilde{t},t)}\ I_{i,j}(t-\tilde{t},t) \text{, as } l_{i,j}(t-\tilde{t},t) \le 1 \nonumber
\end{align}

\noindent By repeatedly applying Lemma \ref{lemma:upper_bound_probability_drift}, we get
\begin{align}
p_{i,j}(t) 
&\le \left(1 + \frac{1}{d}\right) p_{i,j}(t-1) \nonumber \\
&\le \left(1 + \frac{1}{d}\right)^2 p_{i,j}(t-2) \nonumber \\
&\le \left(1 + \frac{1}{d}\right)^{\tilde{t}} p_{i,j}(t-\tilde{t}) \nonumber
\end{align}

\noindent Thus, we can say that
\begin{align}
p_{i,j}(t)\ \widehat{l_{i,j}}(t) 
&\le \frac{1}{d+1} \sum\limits_{\tilde{t}=0}^{d} \left(1 + \frac{1}{d}\right)^{\tilde{t}} \ \frac{p_{i,j}(t-\tilde{t})}{q_{i,j}(t-\tilde{t},t)}\ I_{i,j}(t-\tilde{t},t) \nonumber \\
&\le \frac{e}{d+1} \sum\limits_{\tilde{t}=0}^{d} \frac{p_{i,j}(t-\tilde{t})\ I_{i,j}(t-\tilde{t},t)}{q_{i,j}(t-\tilde{t},t)} \text{, as } e \text{ approximates } \left(1 + \frac{1}{d}\right)^{\tilde{t}} \nonumber 
\end{align}

\noindent Combining this with (\ref{proof:lemma4_2}), we get
\begin{align}
p_{i,j}(t+1)-p_{i,j}(t) 
&\ge - \frac{e \eta}{d+1} \sum\limits_{\tilde{t}=0}^{d} \frac{p_{i,j}(t-\tilde{t})\ I_{i,j}(t-\tilde{t},t)}{q_{i,j}(t-\tilde{t},t)} \nonumber
\end{align}

\noindent which concludes the proof.
\end{IEEEproof}

\BlankLine
\noindent We now proceed with the proof for regret bound.
\begin{IEEEproof}[\textbf{Proof of regret bound}]
\setcounter{equation}{0} % reset numbering from previous appendix
\noindent 
We follow the proof for upper bound on regret for EXP3 \cite{auer2002nonstochastic} and proofs given in \cite{cesa2016delay,alon2017nonstochastic}. Let $W_t = w_{1,j}(t) + \cdots + w_{k,j}(t)$. We try to find a bound on the ratio of weights from one round to the next, i.e., $\dfrac{W_{t+1}}{W_t}$.

\begin{align}
\frac{W_{t+1}}{W_t} &= \sum_{i=1}^{k} \frac{w_{i,j}(t + 1)}{W_t}\textnormal{, given that }W_{t+1} = \sum_{i=1}^{k} w_{i,j}(t + 1) \nonumber \\
&= \sum_{i=1}^{k} \frac{1}{W_t} \cdot \frac{w_{i,j}(t)\ exp(-\eta\ \widehat{l_{i,j}}(t))}{\underset{m \in \mathcal{K}}{\max}\{w_{m,j}(t)\ exp(-\eta\ \widehat{l_{m,j}}(t))\}} \textnormal{, using the weight update rule} \nonumber
\end{align}

\noindent Thus, 
\begin{align}
\underset{m \in \mathcal{K}}{\max}\{w_{m,j}(t)\ exp(-\eta\ \widehat{l_{m,j}}(t))\}\ \frac{W_{t+1}}{W_t} 
&=  \sum_{i=1}^{k} \frac{w_{i,j}(t)}{W_t}\ exp(-\eta\ \widehat{l_{i,j}}(t)) \nonumber \\
&=  \sum_{i=1}^{k} p_{i,j}(t)\ exp(-\eta\ \widehat{l_{i,j}}(t)) \text{, as } \frac{w_{i,j}(t)}{W_t} = p_{i,j}(t)
\label{proof:regret_1}
\end{align}

\noindent From Taylor series, $e^{-x} \le 1 - x + \dfrac{x^2}{2}$, for all $x \ge 0$. \\
\noindent In our case, $x = \eta\ \widehat{l_{i,j}}(t)$. Combining this with (\ref{proof:regret_1}), we get
\begin{align}
\underset{m \in \mathcal{K}}{\max}\{w_{m,j}(t)\ exp(-\eta\ \widehat{l_{m,j}}(t))\} \frac{W_{t+1}}{W_t} 
&\le \sum_{i=1}^{k}  p_{i,j}(t) \ \left[1 - \eta\ \widehat{l_{i,j}}(t) + \frac{\eta^2}{2}\left(\widehat{l_{i,j}}(t)\right)^2\right] \nonumber \\
&\le \sum_{i=1}^{k} p_{i,j}(t) - \eta \sum_{i=1}^{k} p_{i,j}(t)\ \widehat{l_{i,j}}(t) + \frac{\eta^2}{2} \sum_{i=1}^{k} p_{i,j}(t) \left(\widehat{l_{i,j}}(t)\right)^2 \nonumber \\
&\le 1 - \eta \sum_{i=1}^{k} p_{i,j}(t)\ \widehat{l_{i,j}}(t) + \frac{\eta^2}{2} \sum_{i=1}^{k} p_{i,j}(t) \left(\widehat{l_{i,j}}(t)\right)^2 \nonumber 
\end{align}

\noindent Taking logarithms on both sides,
\begin{align}
&\ln\left(\underset{m \in \mathcal{K}}{\max}\{w_{m,j}(t)\ exp(-\eta\ \widehat{l_{m,j}}(t))\} \frac{W_{t+1}}{W_t}\right) \le \ln\left(1 - \eta \sum_{i=1}^{k} p_{i,j}(t) \widehat{l_{i,j}}(t) + \frac{\eta^2}{2} \sum_{i=1}^{k} p_{i,j}(t) \left(\widehat{l_{i,j}}(t)\right)^2\right) 
\label{proof:regret_7}
\end{align}

\noindent $\ln(1 - x) \le -x$ for all $x \ge 0$. In our case, $\displaystyle x = \eta \sum_{i=1}^{k} p_{i,j}(t)\ \widehat{l_{i,j}}(t) - \frac{\eta^2}{2} \sum_{i=1}^{k} p_{i,j}(t) \left(\widehat{l_{i,j}}(t)\right)^2$

\noindent Therefore, combining this with (\ref{proof:regret_7}), we get
\begin{align}
&\ln\left(\underset{m \in \mathcal{K}}{\max}\{w_{m,j}(t)\ exp(-\eta\ \widehat{l_{m,j}}(t))\}\right) + \ln W_{t+1} - \ln W_t 
&\le -\eta \sum_{i=1}^{k} p_{i,j}(t)\ \widehat{l_{i,j}}(t) + \frac{\eta^2}{2} \sum_{i=1}^{k} p_{i,j}(t) \left(\widehat{l_{i,j}}(t)\right)^2\nonumber
\end{align}

\noindent Summing over t,
\begin{align}
&\sum_{t=1}^{T}\left(\ln\left(\underset{m \in \mathcal{K}}{\max}\{w_{m,j}(t)\ exp(-\eta\ \widehat{l_{m,j}}(t))\}\right) + \ln W_{t+1} - \ln W_t\right) 
&\le -\eta \sum_{t=1}^{T} \sum_{i=1}^{k} p_{i,j}(t) \widehat{l_{i,j}}(t) + \dfrac{\eta^2}{2} \sum_{t=1}^{T} \sum_{i=1}^{k} p_{i,j}(t) \left(\widehat{l_{i,j}}(t)\right)^2
\label{proof:regret_8}
\end{align}

\begin{align}
W_{T+1} &\ge w_{i,j}(T+1) \nonumber \\
w_{i,j}(T+1) &= \frac{w_{i,j}(T) \ exp(-\eta\ \widehat{l_{i,j}}(T))}{\underset{m \in \mathcal{K}}{\max}\{w_{m,j}(T)\ exp(-\eta\ \widehat{l_{m,j}}(T))\}} \text{, from the weight update rule}\nonumber \\
&= w_{i,j}(T - 1)\cdot \frac{exp(-\eta\ \widehat{l_{i,j}}(T - 1))}{\underset{m \in \mathcal{K}}{\max}\{w_{m,j}(T-1)\ exp(-\eta\ \widehat{l_{m,j}}(T-1))\}}\cdot \frac{exp(-\eta\ \widehat{l_{i,j}}(T))}{\underset{m \in \mathcal{K}}{\max}\{w_{m,j}(T)\ exp(-\eta\ \widehat{l_{m,j}}(T))\}} \nonumber \\
&=\prod_{t=1}^T \frac{exp(-\eta\ \widehat{l_{i,j}}(t))}{\underset{m \in \mathcal{K}}{\max}\{w_{m,j}(t)\ exp(-\eta\ \widehat{l_{m,j}}(t))\}} \text{, since } w_{i,j}(1) = 1 \nonumber \\
&=\frac{exp\left(\displaystyle\sum_{t=1}^T\ -\eta\ \widehat{l_{i,j}}(t)\right)}{\displaystyle\prod_{t=1}^T \underset{m \in \mathcal{K}}{\max}\{w_{m,j}(t)\ exp(-\eta\ \widehat{l_{m,j}}(t))\}} \nonumber \\
\text{Thus, } W_{T+1} &\ge \frac{exp\left(-\eta \displaystyle\sum_{t=1}^T\ \widehat{l_{i,j}}(t)\right)}{\displaystyle\prod_{t=1}^T \underset{m \in \mathcal{K}}{\max}\{w_{m,j}(t)\ exp(-\eta\ \widehat{l_{m,j}}(t))\}} \nonumber
\end{align}

\noindent Using this to solve the left-hand side of (\ref{proof:regret_8}), in which $\displaystyle\sum_{t=1}^T \left(\ln  W_{t+1} - \ln  W_t\right)$ is a telescoping sum, we get
\begin{align}
&\sum_{t=1}^{T}\left(\ln\left(\underset{m \in \mathcal{K}}{\max}\{w_{m,j}(t)\ exp(-\eta\ \widehat{l_{m,j}}(t))\}\right) + \ln W_{t+1} - \ln W_t\right) \nonumber \\
&\hspace{5 mm} \ge \sum_{t=1}^{T} \ln\left(\underset{m \in \mathcal{K}}{\max}\{w_{m,j}(t)\ exp(-\eta\ \widehat{l_{m,j}}(t))\}\right)  + \ln W_{T+1} - \ln W_1 \nonumber \\
&\hspace{5 mm} \ge \sum_{t=1}^{T} \ln\left(\underset{m \in \mathcal{K}}{\max}\{w_{m,j}(t)\ exp(-\eta\ \widehat{l_{m,j}}(t))\}\right) + \sum_{t=1}^{T}-\eta \widehat{l_{i,j}}(t) - \ln \left(\prod_{t=1}^T \underset{m \in \mathcal{K}}{\max}\{w_{m,j}(t)\ exp(-\eta\ \widehat{l_{m,j}}(t))\right)  - \ln k \nonumber \\
&\hspace{5 mm} \ge \sum_{t=1}^{T} \ln\left(\underset{m \in \mathcal{K}}{\max}\{w_{m,j}(t)\ exp(-\eta\ \widehat{l_{m,j}}(t))\}\right) - \eta \sum_{t=1}^{T} \widehat{l_{i,j}}(t) - \sum_{t=1}^{T} \ln\left(\underset{m \in \mathcal{K}}{\max}\{w_{m,j}(t)\ exp(-\eta\ \widehat{l_{m,j}}(t))\}\right) - \ln k \nonumber \\
&\hspace{5 mm} \ge -\eta \sum_{t=1}^{T} \widehat{l_{i,j}}(t) - \ln k \nonumber 
\end{align}

\noindent Combining this with (\ref{proof:regret_8}), we get
\begin{align}
-\eta \sum_{t=1}^{T} \widehat{l_{i,j}}(t) - \ln k 
&\le -\eta \sum_{t=1}^{T} \sum_{i=1}^{k} p_{i,j}(t) \widehat{l_{i,j}}(t) + \dfrac{\eta^2}{2} \sum_{t=1}^{T} \sum_{i=1}^{k} p_{i,j}(t) \left(\widehat{l_{i,j}}(t)\right)^2 \nonumber
\end{align}

\noindent Multiplying both sides by $\dfrac{1}{\eta}$ and rearranging, we get
\begin{align}
\sum_{t=1}^{T} \sum_{i=1}^{k} p_{i,j}(t) \widehat{l_{i,j}}(t) 
&\le \sum_{t=1}^{T} \widehat{l_{i,j}}(t) + \dfrac{\eta}{2} \sum_{t=1}^{T} \sum_{i=1}^{k} p_{i,j}(t) \left(\widehat{l_{i,j}}(t)\right)^2 + \frac{1}{\eta} \ln k  \nonumber
\end{align}

\noindent Using the loss estimate rule and starting from $t = d + 1$ (for simplicity), we get
\begin{align}
&\frac{1}{d+1} \sum_{t=d+1}^{T} \sum_{i=1}^{k} p_{i,j}(t) \sum\limits_{\tilde{t}=0}^{d} \frac{l_{i,j}(t-\tilde{t}, t)\ I_{i,j}(t-\tilde{t}, t)}{q_{i,j}(t-\tilde{t},t)} \nonumber \\
& \hspace{3 mm} \le \frac{1}{d+1} \sum_{t=d+1}^{T}  \sum\limits_{\tilde{t}=0}^{d} \frac{l_{i,j}(t-\tilde{t},t)\ I_{i,j}(t-\tilde{t},t)}{q_{i,j}(t-\tilde{t},t)} + \frac{\eta}{2} \sum_{t=d+1}^{T} \sum_{i=1}^{k} p_{i,j}(t) \left(\frac{1}{d+1} \sum\limits_{\tilde{t}=0}^{d} \frac{l_{i,j}(t-\tilde{t},t)\ I_{i,j}(t-\tilde{t},t)}{q_{i,j}(t-\tilde{t},t)}\right)^2 + \frac{1}{\eta} \ln k  
\label{proof:regret_9}
\end{align}

\noindent We bound the term on the left-hand side of (\ref{proof:regret_9}) and the second term on its right-hand side separately. We start by lower bounding the term on the left-hand side.
A repeated application of Lemma \ref{lemma:lower_bound_probability_drift}, for $\tilde{t} = 0, \cdots, d$, yields
\begin{align}
p_{i,j}(t) 
&\ge p_{i,j}(t-1) - \frac{e \eta}{d+1} \sum\limits_{\tilde{t}=0}^{d} \frac{p_{i,j}(t-1-\tilde{t})\ I_{i,j}(t-1-\tilde{t},t)}{q_{i,j}(t-1-\tilde{t},t)} \nonumber \\
&\ge p_{i,j}(t-2) -\frac{e \eta}{d+1} \sum\limits_{\tilde{t}=0}^{d} \frac{p_{i,j}(t-2-\tilde{t})\ I_{i,j}(t-2-\tilde{t},t)}{q_{i,j}(t-2-\tilde{t},t)} - \frac{e \eta}{d+1} \sum\limits_{\tilde{t}=0}^{d}  \frac{p_{i,j}(t-1-\tilde{t})\ I_{i,j}(t-1-\tilde{t},t)}{q_{i,j}(t-1-\tilde{t},t)} \nonumber \\
&\ge p_{i,j}(t-\tilde{t}) - \frac{e \eta}{d+1} \sum\limits_{h=1}^{\tilde{t}} \sum\limits_{r=0}^{d} \frac{p_{i,j}(t-h-r)\ I_{i,j}(t-h-r,t)}{q_{i,j}(t-h-r,t)} \nonumber
\end{align}

\noindent Therefore,
\begin{align}
&\frac{1}{d+1}\sum_{t=d+1}^{T} \sum_{i=1}^{k} p_{i,j}(t) \sum\limits_{\tilde{t}=0}^{d} \frac{l_{i,j}(t-\tilde{t},t)\ I_{i,j}(t-\tilde{t},t)}{q_{i,j}(t-\tilde{t},t)} \nonumber \\
&\hspace{5 mm}= \frac{1}{d+1}\sum_{t=d+1}^{T} \sum_{i=1}^{k} \sum\limits_{\tilde{t}=0}^{d} \frac{l_{i,j}(t-\tilde{t},t)\ I_{i,j}(t-\tilde{t},t)}{q_{i,j}(t-\tilde{t},t)} \biggl(p_{i,j}(t-\tilde{t})- \frac{e \eta}{d+1} \sum\limits_{h=1}^{\tilde{t}} \sum\limits_{r=0}^{d}  \frac{p_{i,j}(t-h-r)\ I_{i,j}(t-h-r,t)}{q_{i,j}(t-h-r,t)}\biggr) \nonumber \\
&\hspace{5 mm}= \frac{1}{d+1} \sum_{t=d+1}^{T} \sum_{i=1}^{k} \sum\limits_{\tilde{t}=0}^{d} \frac{p_{i,j}(t-\tilde{t})\ l_{i,j}(t-\tilde{t},t)\ I_{i,j}(t-\tilde{t},t)}{q_{i,j}(t-\tilde{t},t)} \nonumber \\
&\hspace{10 mm} - \frac{e \eta}{(d + 1)^2} \sum_{t=d+1}^{T} \sum_{i=1}^{k} \sum\limits_{\tilde{t}=0}^{d} \frac{l_{i,j}(t-\tilde{t},t)\ I_{i,j}(t-\tilde{t},t)}{q_{i,j}(t-\tilde{t},t)}\ \sum\limits_{h=1}^{\tilde{t}} \sum\limits_{r=0}^{d} \frac{p_{i,j}(t-h-r)\ I_{i,j}(t-h-r,t)}{q_{i,j}(t-h-r,t)}
\label{proof:regret_10}
\end{align}

\noindent We now upper bound the second term on the right-hand side of (\ref{proof:regret_9}). 
\begin{align}
\left(\frac{1}{d+1} \sum\limits_{\tilde{t}=0}^{d} \frac{l_{i,j}(t-\tilde{t},t)\ I_{i,j}(t-\tilde{t},t)}{q_{i,j}(t-\tilde{t},t)}\right)^2 & \le \left(\frac{1}{d+1}  \sum\limits_{\tilde{t}=0}^{d} \frac{I_{i,j}(t-\tilde{t},t)}{q_{i,j}(t-\tilde{t},t)}\right)^2 \text{, given that } l_{i,j}(t-\tilde{t},t) \le 1 \nonumber \\
& \le \frac{1}{d+1} \sum\limits_{\tilde{t}=0}^{d} \frac{I_{i,j}(t-\tilde{t},t)}{\left(q_{i,j}(t-\tilde{t},t)\right)^2} \text{ , using Jensen's inequality} \nonumber
% see Jensen's inequality @ http://sepwww.stanford.edu/sep/prof/pvi/jen/paper_html/node3.html
\end{align}

\noindent We recall, from the poof of Lemma \ref{lemma:lower_bound_probability_drift}, that a repeated application of Lemma \ref{lemma:upper_bound_probability_drift} yields
\begin{align}
p_{i,j}(t) 
&\le \left(1 + \dfrac{1}{d}\right)^{\tilde{t}} p_{i,j}(t - \tilde{t}) \nonumber \\
&\le e\ p_{i,j}(t - \tilde{t}) \text{, given that } e \text{ approximates } \left(1 + \dfrac{1}{d}\right)^{\tilde{t}}\nonumber
\end{align}
\noindent such that
\begin{align}
\frac{\eta}{2} \sum_{t=d+1}^{T} \sum_{i=1}^{k} p_{i,j}(t) \left(\frac{1}{d+1} \sum\limits_{\tilde{t}=0}^{d} \frac{l_{i,j}(t-\tilde{t},t)\ I_{i,j}(t-\tilde{t},t)}{q_{i,j}(t-\tilde{t},t)}\right)^2 
&\le \frac{e \eta}{2(d + 1)} \sum_{t=d+1}^{T} \sum_{i=1}^{k} \sum\limits_{\tilde{t}=0}^{d} \frac{p_{i,j}(t-\tilde{t})\ I_{i,j}(t-\tilde{t},t)}{\left(q_{i,j}(t-\tilde{t},t)\right)^2}
\label{proof:regret_11}
\end{align}

\noindent Combining (\ref{proof:regret_10}) and (\ref{proof:regret_11}) with (\ref{proof:regret_9}) and rearranging, we get
\begin{align}
&\underbrace{\frac{1}{d+1}\sum_{t=d+1}^{T} \sum_{i=1}^{k} \sum\limits_{\tilde{t}=0}^{d} \frac{p_{i,j}(t-\tilde{t})\ l_{i,j}(t-\tilde{t},t)\ I_{i,j}(t-\tilde{t},t)}{q_{i,j}(t-\tilde{t},t)}}_{\text{(I)}} \le \frac{1}{d+1} \underbrace{\sum_{t=d+1}^{T} \sum\limits_{\tilde{t}=0}^{d} \frac{l_{i,j}(t-\tilde{t},t)\ I_{i,j}(t-\tilde{t},t)}{q_{i,j}(t-\tilde{t},t)}}_{\text{(II)}} \nonumber \\
&\hspace{5 mm}+ \underbrace{\frac{e \eta}{(d+1)^2} \sum_{t=d+1}^{T} \sum_{i=1}^{k} \sum\limits_{\tilde{t}=0}^{d} \frac{l_{i,j}(t-\tilde{t},t)\ I_{i,j}(t-\tilde{t},t)}{q_{i,j}(t-\tilde{t},t)}\ \sum\limits_{h=1}^{\tilde{t}} \sum\limits_{r=0}^{d} \frac{p_{i,j}(t-h-r)\ I_{i,j}(t-h-r,t)}{q_{i,j}(t-h-r,t)}}_{\text{(III)}} \nonumber \\
&\hspace{5 mm}+ \underbrace{\frac{e \eta}{2(d+1)} \sum_{t=d+1}^{T} \sum_{i=1}^{k} \sum\limits_{\tilde{t}=0}^{d} \frac{p_{i,j}(t-\tilde{t})\ I_{i,j}(t-\tilde{t},t)}{\left(q_{i,j}(t-\tilde{t},t)\right)^2}}_{\text{(IV)}} + \frac{1}{\eta} \ln k
\label{proof:regret_12}
\end{align}

\noindent We take expectation $\mathbb{E}_{t-\tilde{t}}$ on both sides and solve each term (I to IV) separately. 
% term I
\begin{align}
\mathbb{E}_{t - \tilde{t}}\left[\text{(I)}\right] &= \mathbb{E}_{t-\tilde{t}}\left[\frac{1}{d+1}\sum_{t=d+1}^{T} \sum_{i=1}^{k} \sum\limits_{\tilde{t}=0}^{d} \frac{p_{i,j}(t-\tilde{t})\ l_{i,j}(t-\tilde{t},t)\ I_{i,j}(t-\tilde{t},t)}{q_{i,j}(t-\tilde{t},t)}\right] \nonumber \\
&= \frac{1}{d+1}\sum_{t=d+1}^{T} \sum_{i=1}^{k} \sum\limits_{\tilde{t}=0}^{d} p_{i,j}(t-\tilde{t})\ l_{i,j}(t-\tilde{t},t) \ \mathbb{E}_{t - \tilde{t}}\left[\frac{I_{i,j}(t-\tilde{t},t)}{q_{i,j}(t-\tilde{t},t)}\right] \nonumber \\ 
&= \frac{1}{d+1} \sum_{t=d+1}^{T} \sum_{i=1}^{k} \sum\limits_{\tilde{t}=0}^{d} p_{i,j}(t-\tilde{t})\ l_{i,j}(t-\tilde{t},t) \text{, using Fact \ref{fact:expectation_I_over_q}} \nonumber \\
&= \sum_{t=d+1}^{T} \sum_{i=1}^{k} p_{i,j}(t)l_{i,j}(t) \text{, using Fact \ref{fact:expectation_loss_estimate}} \nonumber \\ 
&=  \sum_{t=1}^{T} \sum_{i=1}^{k} p_{i,j}(t)l_{i,j}(t) - \sum_{t=1}^{d+1} \sum_{i=1}^{k} p_{i,j}(t)l_{i,j}(t) \nonumber \\
&\ge \sum_{t=1}^{T} \sum_{i=1}^{k} p_{i,j}(t)l_{i,j}(t) - d \text{, as } l_{i,j}(t) \le 1 \text{ and } \sum_{i=1}^{k} p_{i,j}(t) = 1
\label{proof:regret_13}
\end{align}

% term II
\begin{align}
\mathbb{E}_{t-\tilde{t}}\left[\text{(II)}\right] &= \mathbb{E}_{t-\tilde{t}}\left[\frac{1}{d+1}\sum_{t=d+1}^{T} \sum\limits_{\tilde{t}=0}^{d}  \frac{l_{i,j}(t-\tilde{t},t)\ I_{i,j}(t-\tilde{t},t)}{q_{i,j}(t-\tilde{t},t)}\right] \nonumber \\
&= \frac{1}{d+1} \sum_{t=d+1}^{T} \sum\limits_{\tilde{t}=0}^{d} l_{i,j}(t-\tilde{t},t)\ \mathbb{E}_{t-\tilde{t}}\left[\frac{I_{i,j}(t-\tilde{t},t)}{q_{i,j}(t-\tilde{t},t)}\right] \nonumber \\
&= \frac{1}{d+1} \sum_{t=d+1}^{T} \sum\limits_{\tilde{t}=0}^{d}  l_{i,j}(t-\tilde{t},t) \text{, using Fact \ref{fact:expectation_I_over_q}} \nonumber \\
%&= \sum_{t=d+1}^{T} l_{i,j}(t) \text{, using Fact \ref{fact:expectation_loss_estimate}}  \nonumber \\
%&\le \sum_{t=1}^{T} l_{i,j}(t)
&= \sum_{t=d+1}^{T} l_{i,j}(t) 
\le \sum_{t=1}^{T} l_{i,j}(t)
\label{proof:regret_14}
\end{align}

% term III
\begin{align}
\mathbb{E}_{t-\tilde{t}}\left[\text{(III)}\right] &= \mathbb{E}_{t-\tilde{t}}\left[\frac{e \eta}{(d+1)^2} \sum_{t=d+1}^{T} \sum_{i=1}^{k} \sum\limits_{\tilde{t}=0}^{d}  \frac{l_{i,j}(t-\tilde{t},t)\ I_{i,j}(t-\tilde{t},t)}{q_{i,j}(t-\tilde{t},t)}\ \sum\limits_{h=1}^{\tilde{t}} \sum\limits_{r=0}^{d} \frac{p_{i,j}(t-h-r)\ I_{i,j}(t-h-r,t)}{q_{i,j}(t-h-r,t)}\right] \nonumber \\
&= \mathbb{E}_{t-\tilde{t}}\left[\frac{e \eta}{(d+1)^2} \sum_{t=d+1}^{T} \sum_{i=1}^{k} \sum\limits_{\tilde{t}=0}^{d} \sum\limits_{h=1}^{\tilde{t}} \sum\limits_{r=0}^{d} \frac{l_{i,j}(t-\tilde{t},t)\ I_{i,j}(t-\tilde{t},t)\ p_{i,j}(t-h-r)\ I_{i,j}(t-h-r,t)}{q_{i,j}(t-\tilde{t},t)\ q_{i,j}(t-h-r,t)}\right] \nonumber
\end{align}

\noindent We consider three cases, depending on the values of the indices $\tilde{t}$, $h$ and $r$.

\noindent \textbf{Case 1}: $t - \tilde{t} > t-h-r$ 
\begin{align}
\mathbb{E}&\left[\frac{l_{i,j}(t-\tilde{t},t) \ I_{i,j}(t-\tilde{t},t)\ p_{i,j}(t-h-r)\ I_{i,j}(t-h-r,t)}{q_{i,j}(t-\tilde{t},t)\ q_{i,j}(t-h-r,t)}\right] \nonumber \\
&\hspace{5 mm} = \mathbb{E}\left[\frac{l_{i,j}(t-\tilde{t},t) \ p_{i,j}(t-h-r)\ I_{i,j}(t-h-r,t)}{q_{i,j}(t-h-r,t)}\ \mathbb{E}_{t -\tilde{t}}\left[\frac{I_{i,j}(t-\tilde{t},t)}{q_{i,j}(t-\tilde{t},t)}\right]\right] \nonumber \\
&\hspace{5 mm} = \mathbb{E}\left[\frac{l_{i,j}(t-\tilde{t},t) \ p_{i,j}(t-h-r)\ I_{i,j}(t-h-r,t)}{q_{i,j}(t-h-r,t)}\right] \text{, using Fact \ref{fact:expectation_I_over_q}} \nonumber \\
&\hspace{5 mm} \le \mathbb{E}\left[\frac{p_{i,j}(t-h-r)\ I_{i,j}(t-h-r,t)}{q_{i,j}(t-h-r,t)}\right] \text{, since } l_{i,j}(t-\tilde{t},t) \le 1 \nonumber \\
&\hspace{5 mm} \le \mathbb{E}\left[p_{i,j}(t-h-r)\ \mathbb{E}_{t-h-r}\left[\frac{I_{i,j}(t-h-r,t)}{q_{i,j}(t-h-r,t)}\right]\right] \nonumber \\
&\hspace{5 mm} \le \mathbb{E}\left[p_{i,j}(t-h-r)\right] \text{, using Fact  \ref{fact:expectation_I_over_q}}\nonumber
\end{align}

\noindent \textbf{Case 2}: $t - \tilde{t} < t-h-r$ 
\begin{align}
\mathbb{E}&\left[\frac{l_{i,j}(t-\tilde{t},t) \ I_{i,j}(t-\tilde{t},t)\ p_{i,j}(t-h-r)\ I_{i,j}(t-h-r,t)}{q_{i,j}(t-\tilde{t},t)\ q_{i,j}(t-h-r,t)}\right] \nonumber \\
&\hspace{5 mm} =  \mathbb{E}\left[\frac{l_{i,j}(t-\tilde{t},t) \ I_{i,j}(t-\tilde{t},t)\ p_{i,j}(t-h-r)}{q_{i,j}(t-\tilde{t},t)}\ \mathbb{E}_{t-h-r}\left[\frac{I_{i,j}(t-h-r,t)}{q_{i,j}(t-h-r,t)}\right]\right] \nonumber \\
&\hspace{5 mm} =  \mathbb{E}\left[\frac{l_{i,j}(t-\tilde{t},t) \ I_{i,j}(t-\tilde{t},t)\ p_{i,j}(t-h-r)}{q_{i,j}(t-\tilde{t},t)}\right] \text{, using Fact \ref{fact:expectation_I_over_q}} \nonumber \\
& \hspace{5 mm} \le  \mathbb{E}\left[\frac{I_{i,j}(t-\tilde{t},t)\ p_{i,j}(t-h-r)}{q_{i,j}(t-\tilde{t},t)}\right] \text{, as } l_{i,j}(t-\tilde{t},t) \le 1 \nonumber \\
&\hspace{5 mm} \le \left(1 + \frac{1}{d}\right)^{\tilde{t}-h-r} \ \mathbb{E}\left[\frac{I_{i,j}(t-\tilde{t},t)\ p_{i,j}(t-\tilde{t})}{q_{i,j}(t-\tilde{t},t)}\right] \text{, by repeatedly applying Lemma \ref{lemma:upper_bound_probability_drift}} \nonumber \\
&\hspace{5 mm} \le  e\ \mathbb{E}\left[p_{i,j}(t-\tilde{t})\ \mathbb{E}_{t-\tilde{t}}\left[\frac{I_{i,j}(t-\tilde{t},t)}{q_{i,j}(t-\tilde{t},t)}\right] \right] \text{, as } e \text{ approximates } \left(1 + \frac{1}{d}\right)^{\tilde{t}-h-r} \nonumber \\
&\hspace{5 mm} \le  e\ \mathbb{E}\left[p_{i,j}(t-\tilde{t})\right]  \text{, using Fact  \ref{fact:expectation_I_over_q}}  \nonumber 
\end{align}

\noindent \textbf{Case 3}: $t - \tilde{t} = t-h-r$. Thus, $p_{i,j}(t-h-r) = p_{i,j}(t-\tilde{t})$ and, given that $h \ge 1$, $\tilde{t} > r$ and $I_{i,j}(t-h-r,t)\ I_{i,j}(t-\tilde{t},t) = I_{i,j}(t-h-r,t)$.
\begin{align}
\mathbb{E}&\left[\frac{l_{i,j}(t-\tilde{t},t) \ I_{i,j}(t-\tilde{t},t)\ p_{i,j}(t-h-r)\ I_{i,j}(t-h-r,t)}{q_{i,j}(t-\tilde{t},t)\ q_{i,j}(t-h-r,t)}\right] \nonumber \\
&\hspace{5 mm} = \mathbb{E}\left[\frac{l_{i,j}(t-\tilde{t},t) \ p_{i,j}(t-h-r)\ I_{i,j}(t-h-r,t)}{q_{i,j}(t-\tilde{t},t)\ q_{i,j}(t-h-r,t)}\right] \nonumber \\
&\hspace{5 mm} = \mathbb{E}\left[\frac{l_{i,j}(t-\tilde{t},t) \ p_{i,j}(t-h-r)}{q_{i,j}(t-\tilde{t},t)}\ \mathbb{E}_{t-h-r}\left[\frac{I_{i,j}(t-h-r,t)}{q_{i,j}(t-h-r,t)}\right]\right] \nonumber \\
&\hspace{5 mm} = \mathbb{E}\left[\frac{l_{i,j}(t-\tilde{t},t) \ p_{i,j}(t-h-r)}{q_{i,j}(t-\tilde{t},t)}\right] \text{, using Fact \ref{fact:expectation_I_over_q}}  \nonumber \\
&\hspace{5 mm} \le \mathbb{E}\left[\frac{p_{i,j}(t-h-r)}{q_{i,j}(t-\tilde{t},t)}\right]  \text{, since } l_{i,j}(t-\tilde{t},t) \le 1  \nonumber \\
&\hspace{5 mm} \le \mathbb{E}\left[\frac{p_{i,j}(t-\tilde{t})}{q_{i,j}(t-\tilde{t},t)}\right] \nonumber
\end{align}

\noindent Considering all the three cases, putting everything together and overapproximating, we get
\begin{align}
&\mathbb{E}_{t-\tilde{t}}\left[\text{(III)}\right] \nonumber \\
&\hspace{1 mm}\le 
\mathbb{E}\left[\frac{e^2 \eta}{(d+1)^2} \sum_{t=d+1}^{T} \sum_{i=1}^{k} \left( \sum_{\tilde{t},h,r: \tilde{t}<h+r} p_{i,j}(t-h-r) + \sum_{\tilde{t},h,r: \tilde{t}>h+r} p_{i,j}(t-\tilde{t}) + \sum_{\tilde{t},h,r: \tilde{t}=h+r} \frac{p_{i,j}(t-\tilde{t})}{q_{i,j}(t-\tilde{t},t)}\right)\right] \nonumber \\
&\hspace{1 mm}\le 
\frac{e^2 \eta}{(d+1)^2}\ \mathbb{E}\left[ \sum_{t=d+1}^{T} \left( \sum_{\tilde{t},h,r: \tilde{t}<h+r} \sum_{i=1}^{k} p_{i,j}(t-h-r) + \sum_{\tilde{t},h,r: \tilde{t}>h+r} \sum_{i=1}^{k} p_{i,j}(t-\tilde{t}) + \sum_{\tilde{t},h,r: \tilde{t}=h+r} \sum_{i=1}^{k} \frac{p_{i,j}(t-\tilde{t})}{q_{i,j}(t-\tilde{t},t)}\right)\right] \nonumber \\
&\hspace{1 mm}\le 
\frac{e^2 \eta}{(d+1)^2}\ \mathbb{E}\left[ \sum_{t=d+1}^{T} \left( \sum_{\tilde{t},h,r: \tilde{t}\neq h+r} \sum_{i=1}^{k} p_{i,j}(t-\tilde{t}) + \sum_{\tilde{t},h,r: \tilde{t}=h+r} \sum_{i=1}^{k} \frac{p_{i,j}(t-\tilde{t})}{q_{i,j}(t-\tilde{t},t)}\right)\right] \nonumber \\
&\hspace{1 mm}\le 
\frac{e^2 \eta}{(d+1)^2}\ \mathbb{E}\left[ \sum_{t=d+1}^{T} \left( \sum_{\tilde{t},h,r: \tilde{t}\neq h+r} \sum_{i=1}^{k} \frac{p_{i,j}(t-\tilde{t})}{q_{i,j}(t-\tilde{t},t)} + \sum_{\tilde{t},h,r: \tilde{t}=h+r} \sum_{i=1}^{k} \frac{p_{i,j}(t-\tilde{t})}{q_{i,j}(t-\tilde{t},t)}\right)\right] \text{, as } q_{i,j}(t-\tilde{t},t) \le 1 \nonumber \\ 
&\hspace{1 mm}\le 
\frac{e^2 \eta}{(d+1)^2}\ \sum_{t=d+1}^{T} \sum_{\tilde{t},h,r} \mathbb{E}\left[\sum_{i=1}^{k} \frac{p_{i,j}(t-\tilde{t})}{q_{i,j}(t-\tilde{t},t)} \right] \nonumber \\
&\hspace{1 mm}\le 
\frac{e^2 \eta}{(d+1)^2}\ \sum_{t=d+1}^{T} \sum_{\tilde{t},h,r} \frac{1}{\max\{\frac{1}{k},b_0\}} \text{ , using Lemma \ref{lemma:sum_networks_p_q}} \nonumber \\
&\hspace{1 mm}\le 
e^2 \eta \sum_{t=d+1}^{T} \sum_{\tilde{t} = 0}^{d} \frac{1}{d+1} \left( \frac{1}{\max\{\frac{1}{k},b_0\}} \right) \sum_{h = 1}^{\tilde{t}} \sum_{r = 0}^{d} \frac{1}{d+1} \nonumber \\
&\hspace{1 mm}\le 
e^2 \eta \sum_{t=d+1}^{T} \sum_{\tilde{t} = 0}^{d} \frac{\tilde{t}}{d+1} \left( \frac{1}{\max\{\frac{1}{k},b_0\}} \right) \text{, as } \sum_{r = 0}^{d} \frac{1}{d+1} = 1 \text{, and } \sum_{h = 1}^{\tilde{t}} 1 = \tilde{t} \nonumber \\
&\hspace{1 mm}\le 
e^2 \eta \sum_{t=1}^{T} \sum_{\tilde{t} = 0}^{d} \frac{\tilde{t}}{d+1} \left( \frac{1}{\max\{\frac{1}{k},b_0\}} \right) \nonumber \\
&\hspace{1 mm} \le 
 \frac{e^2 \eta d T}{2 \max\{\frac{1}{k}, b_0\}} \text{, as } \sum{\tilde{t}=1}^{d} \tilde{t} = \frac{d}{2}
\label{proof:regret_15}
\end{align}

% term IV
\begin{align}
\mathbb{E}_{t-\tilde{t}}\left[\text{(IV)}\right] 
&= \mathbb{E}_{t-\tilde{t}}\left[\frac{e \eta}{2 (d+1)} \sum_{t=d+1}^{T} \sum_{i=1}^{k} \sum\limits_{\tilde{t}=0}^{d} \frac{p_{i,j}(t-\tilde{t})\ I_{i,j}(t-\tilde{t},t)}{\left(q_{i,j}(t-\tilde{t},t)\right)^2}\right] \nonumber \\
&= \frac{e \eta}{2 (d+1)} \sum_{t=d+1}^{T} \sum\limits_{\tilde{t}=0}^{d} \mathbb{E}_{t-\tilde{t}} \left[ \sum_{i=1}^{k} \frac{p_{i,j}(t-\tilde{t})\ I_{i,j}(t-\tilde{t},t)}{\left(q_{i,j}(t-\tilde{t},t)\right)^2}\right] \nonumber \\
&= \frac{e \eta}{2 (d + 1)} \sum_{t=d+1}^{T} \sum\limits_{\tilde{t}=0}^{d} \mathbb{E}_{t-\tilde{t}} \left[ \sum_{i=1}^{k} \frac{p_{i,j}(t-\tilde{t})}{q_{i,j}(t-\tilde{t},t)}\right]  \text{, using Fact \ref{fact:expectation_I_over_q}} \nonumber \\
&= \frac{e \eta}{2 (d+1)} \sum_{t=d+1}^{T} \sum\limits_{\tilde{t}=0}^{d} \left(\frac{1}{\max\{\frac{1}{k},b_0\}}\right) \text{, using Lemma \ref{lemma:sum_networks_p_q}} \nonumber \\
&\le \frac{e \eta}{2 \left(d+1\right)} \sum_{t=1}^{T} \sum\limits_{\tilde{t}=0}^{d} \frac{1}{\max\{\frac{1}{k},b_0\}} \nonumber \\
% &\color{blue}\le \frac{e \eta}{2 \left(d+1\right)} \sum_{t=1}^{T} \sum\limits_{\tilde{t}=0}^{d} \frac{\tilde{t}}{1 - (1- \frac{1}{k})\left(1 - \min\{1 - e^{-1}, b_{\tilde{t}}\}\right)} \nonumber \\
&\le \frac{e \eta T}{2 \max\{\frac{1}{k},b_0\}} 
\label{proof:regret_16} 
\end{align}

\noindent Combining (\ref{proof:regret_13}), (\ref{proof:regret_14}), (\ref{proof:regret_15}) and (\ref{proof:regret_16}) in (\ref{proof:regret_12}), we get
\begin{align}
\sum_{t=1}^{T} \sum_{i=1}^{k} p_{i,j}(t)l_{i,j}(t) 
&\le \sum_{t=1}^{T} l_{i,j}(t) + 
\frac{e^2 \eta  d T}{2\max\{\frac{1}{k},b_0\}} + \frac{e \eta T}{2\max\{\frac{1}{k},b_0\}} + \frac{1}{\eta} \ln k + d \nonumber 
\end{align}

\noindent Let $L_{Co-Bandit}(T) = \sum_{t=1}^{T} \sum_{i=1}^{k} p_{i,j}(t)l_{i,j}(t)$, i.e. denote the expected aggregate loss of C-Bandit, and $L_{min}(T) = \sum_{t=1}^{T} l_{i,j}(t)$, i.e. be the aggregate loss of the best expert. Therefore, overapproximating and simplifying, we get
\begin{align}
\mathbb{E}\left[L_{Co-Bandit}(T) - L_{min}(T)\right] &
\le 
\frac{e^2 (d + 1) \eta T}{2 \max\{\frac{1}{k}, b_0\}} + \frac{1}{\eta} \ln k + d \nonumber 
\end{align}

\noindent We set the following value for $\eta$.
\begin{align}
\eta =
\sqrt{\frac{\max\{\frac{1}{k}, b_0\} \ln k}{e^2 (d + 1) T}} \nonumber
\end{align}

\noindent In Lemma \ref{lemma:upper_bound_probability_drift}, we assumed that $\eta \le \dfrac{1}{ke(d+1)}$. Therefore,
\begin{align}
\sqrt{\frac{\max\{\frac{1}{k}, b_0\} \ln k}{e^2 (d + 1) T}} &\le \dfrac{1}{ke(d+1)} \nonumber \\
\frac{\max\{\frac{1}{k}, b_0\} \ln k}{e^2 (d + 1) T} &\le \dfrac{1}{k^2 e^2 (d+1)^2} \nonumber
\end{align}

\begin{align}
T 
&\le k^2 (d + 1) \max\{\frac{1}{k},b_0\} \ln k \nonumber \\
&\le k (d + 1) \ln k \nonumber
\end{align}

\noindent Hence, we get
\begin{align}
\mathbb{E}\left[L_{Co-Bandit}(T) - L_{min}(T)\right] 
&\le \frac{e}{2} \sqrt{\frac{(d+1) T \ln K}{\max\{\frac{1}{k}, b_0\}}} + e \sqrt{\frac{(d+1) T \ln K}{\max\{\frac{1}{k}, b_0\}}} + d \nonumber \\
&\le 2 e \sqrt{\frac{(d+1) T \ln K}{\max\{\frac{1}{k}, b_0\}}} + d \nonumber
\end{align}
\noindent which concludes the proof.
\end{IEEEproof}
\section{Proof of convergence} \label{appendix:convergence}
    % \begin{IEEEproof}[\textbf{Proof}]
%\textbf{\textit{Proof:}}
\setcounter{equation}{0} % reset numbering from previous appendix
\noindent 
We follow the proofs given in \cite{kleinberg2009multiplicative} and \cite{tekin2011performance}. 
We analyze the evolution of the probability of network $i$. In addition to a random choice from the probability distribution, a device also explores network(s) unheard of for at least $x$ time slots with probability $\frac{1}{n}$. We assume that $n$ tends to $\infty$; hence $\frac{1}{n}$ tends to zero.\\

% \noindent Hence, we can define the probability of a network $i$ for device $j$ as 
% \begin{align}
% p_{i,j}(t) 
% &\le \left(1 - \gamma\right) \frac{w_{i,j}(t)}{\sum_{m=1}^{k} w_{m,j}(t)} + \frac{\gamma}{k - 1} \text{, where } \gamma = \frac{k-1}{n} \nonumber 
% % &\le \left(1 - \gamma\right) \frac{w_{i,j}(t)}{\sum_{m=1}^{k} w_{m,j}(t)} + \frac{1}{n} \nonumber
% \end{align}

\noindent We consider two cases, namely (1) when the loss of network $i$ has been observed by device $j$ or learnt from its neighbors, and (2) when the loss of network $i$ is unknown to device $j$. The former case occurs with probability $q$ and the latter with probability $1-q$, as defined in Algorithm \ref{algorithm:collaborativeEWA}.  

\noindent We start with the first case, i.e. when device $j$ has observed the loss of network $i$ or has learnt about it from its neighbors.

\begin{align}
p_{i,j}(t + 1) 
&= \dfrac{w_{i,j}(t+1)}{\sum\limits_{m=1}^{k} w_{m,j}(t+1)}  \text{, from the probability update rule} \nonumber \\
&= \dfrac{\dfrac{w_{i,j}(t)\ exp(-\eta\ \widehat{l_{i,j}}(t))}{\underset{y \in \mathcal{K}}{\max}\{w_{y,j}(t)\ exp(-\eta\ \widehat{l_{y,j}}(t))\}}}{\sum\limits_{m=1}^{k} \dfrac{w_{m,j}(t)\ exp(-\eta\ \widehat{l_{m,j}}(t))}{\underset{y \in \mathcal{K}}{\max}\{w_{y,j}(t)\ exp(-\eta\ \widehat{l_{y,j}}(t))\}}}  \text{, using the weight update rule} \nonumber \\
&=\dfrac{w_{i,j}(t)\ exp(-\eta\ \widehat{l_{i,j}}(t))}{\sum\limits_{m=1}^{k} w_{m,j}(t)\ exp(-\eta\ \widehat{l_{m,j}}(t))}  
\label{proof:convergence_1}
\end{align}

\noindent From the probability update rule, we have
\begin{align}
p_{i,j}(t) 
&= \dfrac{w_{i,j}(t)}{\sum\limits_{m=1}^{k} w_{m,j}(t)} \nonumber \\
w_{i,j}(t) 
&= \sum\limits_{m=1}^{k} w_{m,j}(t)\ p_{i,j}(t) 
\label{proof:convergence_2}
\end{align}

\noindent Combining (\ref{proof:convergence_2}) with (\ref{proof:convergence_1}), we get
\begin{align}
p_{i,j}(t + 1) 
&= \dfrac{\sum\limits_{m=1}^{k} w_{m,j}(t) \ p_{i,j}(t)  \ exp(-\eta\ \widehat{l_{i,j}}(t))}{\sum\limits_{m=1}^{k} \sum\limits_{y=1}^{k} w_{y,j}(t) \ p_{m,j}(t) \ exp(-\eta\ \widehat{l_{m,j}}(t))}  \nonumber \\
&= \dfrac{p_{i,j}(t) \ exp(-\eta\ \widehat{l_{i,j}}(t))}{\sum\limits_{m=1}^{k} p_{m,j}(t) \ exp(-\eta\ \widehat{l_{m,j}}(t))} \nonumber
\end{align}

\noindent We drop the discrete time script $t$ and compute the continuous time process by deriving the limit of the probability update rule as $\eta \rightarrow 0$, a first order differential equation known as the replicator dynamic. This is consistent with Lemma \ref{lemma:upper_bound_probability_drift} (in Appendix \ref{appendix:regretBound}) and Theorem \ref{theorem:regretBound} in which we assumed arbitrarily small values of $\eta$.

\begin{align}
p_{i,j}
&= \dfrac{p_{i,j}\ exp\left(-\eta\ \widehat{l_{i,j}}\right)}{\sum\limits_{m=1}^{k} p_{m,j} \ exp\left(-\eta\ \widehat{l_{m,j}}\right)}   \nonumber 
\end{align}

\begin{align}
\dot{p_{i,j}}
&= \lim_{\eta \rightarrow 0}\ \frac{dp_{i,j}}{d\eta} \nonumber \\
&= \lim_{\eta \rightarrow 0}\ \frac{d}{d\eta}\left(\dfrac{p_{i,j}  \ exp\left(-\eta\ \widehat{l_{i,j}}\right)}{\sum\limits_{m=1}^{k} \ p_{m,j} \ exp\left(-\eta\ \widehat{l_{m,j}}\right)} \right) \nonumber \\
&= \lim_{\eta \rightarrow 0} \left(\frac{-\sum\limits_{m=1}^{k} p_{m,j} \ exp\left(-\eta\ \widehat{l_{m,j}}\right) \ p_{i,j}  \ exp\left(-\eta\ \widehat{l_{i,j}}\right) \widehat{l_{i,j}}  + \ p_{i,j}  \ exp\left(-\eta\ \widehat{l_{i,j}}\right) \sum\limits_{m=1}^{k} p_{m,j} \ exp\left(-\eta\ \widehat{l_{m,j}}\right) \widehat{l_{m,j}}}{\left(\sum\limits_{m=1}^{k} p_{m,j} \ exp\left(-\eta\ \widehat{l_{m,j}}\right)\right)^{2}}\right) \nonumber \\
&= \frac{- p_{i,j} \ \widehat{l_{i,j}} \sum\limits_{m=1}^{k} p_{m,j} +  p_{i,j} \sum\limits_{m=1}^{k} p_{m,j} \ \widehat{l_{m,j}}}{\left(\sum\limits_{m=1}^{k} p_{m,j}\right)^{2}} \nonumber \\
&= p_{i,j} \sum\limits_{m=1}^{k} p_{m,j} \ \left(\widehat{l_{m,j}} - \widehat{l_{i,j}}\right) \label{proof:convergence_3}
\end{align}

\noindent We now consider the second case, i.e. when the loss of network $i$ is unknown to device $j$; we assume here that the loss of the other networks are known to device $j$.

\begin{align}
p_{i,j}(t + 1) 
&= \dfrac{w_{i,j}(t+1)}{\sum\limits_{m=1}^{k} w_{m,j}(t+1)} \text{, from the probability update rule} \nonumber \\
&= \dfrac{\dfrac{w_{i,j}(t)}{\underset{y \in \mathcal{K}}{\max}\{w_{y,j}(t)\ exp(-\eta\ \widehat{l_{y,j}}(t))\}}}{\sum\limits_{m \in \mathcal{K} - \{i\}} \dfrac{w_{m,j}(t)\ exp(-\eta\ \widehat{l_{m,j}}(t))}{\underset{y \in \mathcal{K}}{\max}\{w_{y,j}(t)\ exp(-\eta\ \widehat{l_{y,j}}(t))\}} + \dfrac{w_{i,j}(t)}{\underset{y \in \mathcal{K}}{\max}\{w_{y,j}(t)\ exp(-\eta\ \widehat{l_{y,j}}(t))\}}}  \text{, using the weight update rule} \nonumber \\
&=\dfrac{w_{i,j}(t)}{\sum\limits_{m\in\mathcal{K} - \{i\}} w_{m,j}(t)\ exp(-\eta\ \widehat{l_{m,j}}(t)) + w_{i,j}(t)}   
\label{proof:convergence_4}
\end{align}

\noindent Combining (\ref{proof:convergence_2}) with (\ref{proof:convergence_4}), we get
\begin{align}
p_{i,j}(t + 1) 
&= \dfrac{\sum\limits_{m=1}^{k} w_{m,j}(t) \ p_{i,j}(t)  }{\sum\limits_{m\in \mathcal{K} - \{j\}} \sum\limits_{y=1}^{k} w_{y,j}(t) \ p_{m,j}(t) \ exp(-\eta\ \widehat{l_{m,j}}(t)) + \sum\limits_{m=1}^{k} w_{m,j}(t) \ p_{i,j}(t)}   \nonumber \\
&= \dfrac{p_{i,j}(t) }{\sum\limits_{m \in \mathcal{K} - \{i\}} p_{m,j}(t) \ exp(-\eta\ \widehat{l_{m,j}}(t)) + p_{i,j}(t)}  \nonumber
\end{align}

\noindent As for the first case, we drop the discrete time script $t$ and derive the limit of the probability update rule as $\eta \rightarrow 0$.

\begin{align}
p_{i,j}
&= \dfrac{p_{i,j}}{\sum\limits_{m \in \mathcal{K} - \{i\}} p_{m,j} \ exp(-\eta\ \widehat{l_{m,j}}) + p_{i,j}}  \nonumber
\end{align}

\begin{align}
\dot{p_{i,j}}
&= \lim_{\eta \rightarrow 0}\ \frac{dp_{i,j}}{d\eta} \nonumber \\
&= \lim_{\eta \rightarrow 0}\ \frac{d}{d\eta}\left(\dfrac{p_{i,j}}{\sum\limits_{m \in \mathcal{K} - \{i\}} p_{m,j} \ exp(-\eta\ \widehat{l_{m,j}}) + p_{i,j} } \right) \nonumber \\
&= \lim_{\eta \rightarrow 0} \left(\frac{0 - p_{i,j}  \left(-\sum\limits_{m \in \mathcal{K} - \{i\}} p_{m,j} \ exp(-\eta\ \widehat{l_{m,j}})\ \widehat{l_{m,j}}\right)}{\left(\sum\limits_{m \in \mathcal{K} - \{i\}} p_{m,j} \ exp(-\eta\ \widehat{l_{m,j}}) + p_{i,j} \right)^2}\right) \nonumber \\
&= \dfrac{p_{i,j}  \sum\limits_{m \in \mathcal{K} - \{i\}} p_{m,j} \ \widehat{l_{m,j}}}{\left(\sum\limits_{m \in \mathcal{K} - \{i\}} p_{m,j} + p_{i,j}\right)^2}
\nonumber \\
&= p_{i,j} \sum\limits_{m \in \mathcal{K} - \{i\}} p_{m,j}\ \widehat{l_{m,j}} 
\label{proof:convergence_5}
\end{align}

\noindent Then, from (\ref{proof:convergence_3}) and (\ref{proof:convergence_5}), the expected change in $p_{i,j}$ with respect to the probability distribution of device $j$ over all networks $i \in \mathcal{K}$ is given as
\begin{align}
\bar{p}_{i,j} 
&= \mathbb{E}_{j}\left[\dot{p_{i,j}}\right] \nonumber \\
&= \mathbb{E}_{j}\left[\dot{p_{i,j}}\ | \ \widehat{l_{i,j}} \text{ is known}\right] + \mathbb{E}_{j}\left[\dot{p_{i,j}}\ | \ \widehat{l_{i,j}} \text{ is unknown}\right] \nonumber \\
&= q\ \mathbb{E}\left[p_{i,j} \sum\limits_{m=1}^{k} p_{m,j}  \left(\widehat{l_{m,j}} - \widehat{l_{i,j}}\right)\right] + \left(1-q\right) \mathbb{E}\left[p_{i,j} \sum\limits_{m \in \mathcal{K} - \{i\}} p_{m,j}\ \widehat{l_{m,j}}\right] \nonumber \\
&= q\ p_{i,j} \sum\limits_{m=1}^{k} p_{m,j}  \left(l_{m,j} - l_{i,j}\right) + \left(1-q\right) p_{i,j} \sum\limits_{m \in \mathcal{K} - \{i\}} p_{m,j}\ l_{m,j} \nonumber \\
&= q\ p_{i,j} \left( \sum\limits_{m \in \mathcal{K} - \{i\}} p_{m,j} \left(l_{m,j} - l_{i,j}\right)  + p_{i,j} \left(l_{i, j} - l_{i, j}\right) \right) + \left(1-q\right) p_{i,j} \sum\limits_{m \in \mathcal{K} - \{i\}} p_{m,j}\ l_{m,j} \nonumber\\ 
&= q\ p_{i,j} \sum\limits_{m \in \mathcal{K} - \{i\}} p_{m,j} \left(l_{m,j} - l_{i,j}\right) + \left(1-q\right) p_{i,j} \sum\limits_{m \in \mathcal{K} - \{i\}} p_{m,j}\ l_{m,j} \nonumber\\
&= q\ p_{i,j} \sum\limits_{m \in \mathcal{K} - \{i\}} p_{m,j}\ l_{m,j} - q\ p_{i,j} \sum\limits_{m \in \mathcal{K} - \{i\}} p_{m,j}\ l_{i,j} + \left(1-q\right) p_{i,j} \sum\limits_{m \in \mathcal{K} - \{i\}} p_{m,j}\ l_{m,j} \nonumber\\
&= p_{i,j} \sum\limits_{m \in \mathcal{K} - \{i\}} p_{m,j}\ l_{m,j} - q\ p_{i,j} \sum\limits_{m \in \mathcal{K} - \{i\}} p_{m,j}\ l_{i,j} \nonumber \\
&= p_{i,j}\sum\limits_{m \in \mathcal{K} - \{i\}} p_{m,j} \left(l_{m,j} - q\  l_{i,j}\right) \nonumber 
\end{align}

\noindent Taking expectation with respect to other devices' actions,
\begin{align}
\xi_{i,j} 
&= \mathbb{E}_{-j}\left[\bar{p}_{i,j} \right] \nonumber \\
& =p_{i,j} \sum\limits_{m \in \mathcal{K} - \{i\}} p_{m,j} \left(\mathbb{E}_{-j}\left[l_{m,j}\right] - q\ \mathbb{E}_{-j}\left[l_{i,j}\right]\right) \nonumber\\
&= p_{i,j} \sum\limits_{m \in \mathcal{K} - \{i\}} p_{m,j} \left(l_{m,j} - q\ l_{i,j}\right) \nonumber
\end{align}

\noindent Given that this replicator dynamics is the same as the one \cite{kleinberg2009multiplicative}, except that we have a factor $q$ of $l_{i,j}$, the rest of the proof follows from \cite{kleinberg2009multiplicative}.\\

%\noindent In the full information setting, $b = 1$ and $q$ will tend to 1. In that case, we get a replicator dynamic identical to the one in \cite{kleinberg2009multiplicative}. With a drop in the probability of hearing about network $i$, i.e., the value of $q$, the value of $\xi_{i,j}$ will rise, implying a slower convergence. Furthermore, in the bandit setting, $b = 0$ and $q = p_{i,j}$. Given our definition of loss, $\xi_{i,j}$ will be zero all the time. Hence, the algorithm never converges to the right state. \\ 
\end{document}